\documentclass[hyperref]{solarphysics}
\usepackage[hyperref,optionalrh,solaromanenum]{spr-sola-addons} 
\usepackage{graphicx}                    
\usepackage{courier}                    
\usepackage{amssymb, mathrsfs, adjustbox}                    

\usepackage{amsmath}
\usepackage{color}                       
\usepackage{breakurl}                         

\usepackage[top=3cm, left=3cm, right=3cm]{geometry}
\hypersetup{colorlinks=true, allcolors=blue}

\newcommand{\bfv}{\boldsymbol{v}}

\newcommand{\bfk}{\boldsymbol{k}}
\newcommand{\bfb}{\boldsymbol{B}}
\newcommand{\bfg}{\boldsymbol{g}}
\newcommand{\bfA}{\boldsymbol{A}}
\newcommand{\bfkappa}{\boldsymbol{\kappa}}
\newcommand{\HL}{\mathscr{L}}
\newcommand{\bfeb}{\boldsymbol{e}_B}
\newcommand{\bex}{~\boldsymbol{e}_x}
\newcommand{\bey}{~\boldsymbol{e}_y}
\newcommand{\bez}{~\boldsymbol{e}_z}
\newcommand{\amat}{\mathcal{A}}
\newcommand{\bmat}{\mathcal{B}}
\newcommand{\bfX}{\boldsymbol{X}}
\newcommand{\Qi}{\mathcal{Q}_i}
\newcommand{\Qifull}{\rho_0^2\HL_\rho - \left(\kappa_\parallel k_\parallel^2 + \rho_0 \HL_T\right)c_i^2}
\newcommand{\Qai}{\mathcal{Q}_{ai}}
\newcommand{\Qaifull}{\rho_0^2\HL_\rho - \left(\kappa_\parallel k_\parallel^2 + \rho_0 \HL_T\right)\left(c_A^2 + c_i^2\right)}

\begin{document}

\begin{article}

\begin{opening}

\title{Magnetohydrodynamic spectroscopy of a non-adiabatic solar atmosphere}


\author[addressref={aff1},email={niels.claes@kuleuven.be}]{\inits{N.}\fnm{Niels}~\lnm{Claes}\orcid{0000-0002-8720-9119}}
\author[addressref={aff1}, email={rony.keppens@kuleuven.be}]{\inits{R.}\fnm{Rony}~\lnm{Keppens}\orcid{0000-0003-3544-2733}}


\address[id=aff1]{Centre for mathematical Plasma-Astrophysics, KU Leuven, Celestijnenlaan 200B, 3001 Leuven, Belgium}


\begin{abstract}

	The quantification of all possible waves and instabilities in a given system is of paramount importance, and knowledge of the full magnetohydrodynamic (MHD) spectrum allows one to predict the (in)stability of a given equilibrium state.
	This is highly relevant in many (astro)physical disciplines, and when applied to the solar atmosphere it may yield various new insights in processes like prominence formation and coronal loop oscillations.
	In this work we present a detailed, high-resolution spectroscopic study of the solar atmosphere, where we use our newly developed \texttt{Legolas} code to calculate the full spectrum with corresponding eigenfunctions of equilibrium configurations that are based on fully realistic solar atmospheric models, including gravity, optically thin radiative losses and thermal conduction. Special attention is given to thermal instabilities, known to be responsible for the formation of prominences, together with a new outlook on the thermal and slow continua and how they behave in different chromospheric and coronal regions. We show that thermal instabilities are unavoidable in our solar atmospheric models and that there exist certain regions where both the thermal, slow and fast modes all have unstable wave mode solutions. We also encounter regions where the slow and thermal continua become purely imaginary and merge on the imaginary axis. The spectra discussed in this work illustrate clearly that thermal instabilities (both discrete and continuum modes) and magneto-thermal overstable propagating modes are ubiquitous throughout the solar atmosphere, and may well be responsible for much of the observed fine-structuring and multi-thermal dynamics.
\end{abstract}



\end{opening}


\section{Introduction} \label{sect: intro}
The solar atmosphere shows a myriad of thermodynamically fascinating features, from solar prominences to coronal rain, over plumes, jets, etc. Extensive research on all of these features has been done over the past decennia, and currently nonlinear MHD simulations are starting to resolve aspects at resolutions rivaling upcoming and future solar telescopes. E.g. \citet{ruan2020} showed for the first time a fully self-consistent solar flare model achieving resolutions of 50 km, while \citet{jenkins2021} focused more on the physical processes driving prominence formation, increasing the resolution even further down to 5 km. The basic physical process behind prominence formation is thermal instability, first investigated by \citet{parker1953} and demonstrated through direct observations by for example \citet{berger2012}, as well as in numerical simulations \citep{xia2016, claes2020}. Since many linear waves and instabilities can be at play and interact in realistic solar atmospheric evolutions, modern nonlinear simulations can benefit greatly from the full knowledge of all linear instabilities and eigenoscillations of a given configuration. Some early attempts were made by for example \citet{nye1976}, who discussed an analytical model of the solar atmosphere based on an isothermal slab, where certain mode types of the magnetized atmosphere could be computed analytically in ideal MHD. \citet{vanderlinden1991} did pioneering work in linear MHD including non-adiabatic effects, hinting at the role of magneto-thermal modes in prominence fine-structuring, but since then further research into this topic has stalled.

While fully nonlinear, quasi-realistic numerical simulations can accurately model a plethora of physical phenomena, one can not underestimate the importance of linear MHD. Looking at the saturation of a single unstable mode in the context of prominence formation is one thing, but a detailed knowledge of which particular mode among a plethora of unstable eigenmodes is responsible for the evolution can only be obtained through a detailed spectroscopic analysis of a given state. This has been realized by \citet{demaerel2016}, where they showed (in ideal MHD, including self-gravity) that spectral theory actually governs the stability of single-fluid evolving time-dependent plasmas, at any point during their nonlinear evolution. Knowing all MHD modes of possibly coupled mode types in the magnetized solar atmosphere, and how they modify as a result of including relevant non-adiabatic effects, is thus a clear necessity. We here set forth to do this for a realistic solar atmosphere model, in which all relevant physical effects are included for thermal instability and prominence formation, that is, external gravity, radiative cooling and anisotropic thermal conduction. To date there exists no detailed MHD spectroscopic treatment of such a self-consistent model.

There are in fact a wide range of existing interesting studies regarding thermal instabilities and waves in non-adiabatic solar atmosphere-like settings. \citet{zavershinskii2019} demonstrated the formation of slow magneto-acoustic wave trains due to linear dispersion associated with heating-cooling misbalance, while recently \citet{duckenfield2021} looked at slow mode damping rates in solar coronal conditions under the influence of thermal imbalance and showed that these may be intricately linked. \citet{ledentsov2021} on the other hand looked at thermal instability in preflare current layers including viscosity and the effect a longitudinal magnetic field has on spatial stabilization. However, studies like these are quite limited, in the sense that they have to drastically reduce their mathematical models to either homogeneous background conditions in 2- or 3-D setups, or 1D treatments along field lines in order to be able to derive a dispersion relation or make it analytically tractable.

Armed with the recently developed \texttt{Legolas} code \citep{legolas_paper_2020}, we are now able to do a complete eigenspectrum analysis on a fully realistic solar atmosphere model where we adopt the widely used semi-empirical temperature profile proposed by \citet{avrett2008}, and include all aforementioned physical effects. We pay special attention to both the thermal and slow continua, both of which play a very important role in the (in)stability of a given equilibrium configuration. The intricate structure of the thermal, slow and Alfv\'en continua, and the way the many discrete modes organize in (coupled) thermal, slow, Alfv\'en and fast wave sequences are demonstrated here for the first time, giving us a linear ``preview" of how nonlinear simulations should develop as a result of (interacting) instabilities. The main advantage of using \texttt{Legolas} here is that we are not bound by the same limitations as previous works, in the sense that we do not need any dispersion relation or continuous background profiles in order to calculate growth rates, meaning we can go far beyond that what has already been done in existing literature in our treatment of the solar atmosphere.

In Section \ref{sec: background} we briefly touch upon the system of equations and how these are treated by \texttt{Legolas}, followed by a detailed discussion of the slow and thermal continua and how these relate to mode stability. We first revisit the analytical work by \citet{nye1976} in Section \ref{sec: analytical} and complement this by looking at non-parallel propagation and the inclusion of non-adiabatic effects, something that was impossible in the original analytical treatment. Lastly, in Section \ref{sec: solar-atmo} we look at a fully realistic solar atmosphere model and perform a detailed spectroscopic analysis, with special attention to thermal instabilities.

We demonstrate that thermal instability is ubiquitously present in both the solar chromosphere and corona, and provide a fresh outlook on mode behavior in these regions. In the solar chromosphere, we show that it becomes possible for the slow continua to merge with the imaginary axis, indicating upwards and downwards branches which in turn may become unstable if the conditions are right.

\section{Background} \label{sec: background}
\begin{figure}[t]
	\centering
	\includegraphics[width=0.7\columnwidth]{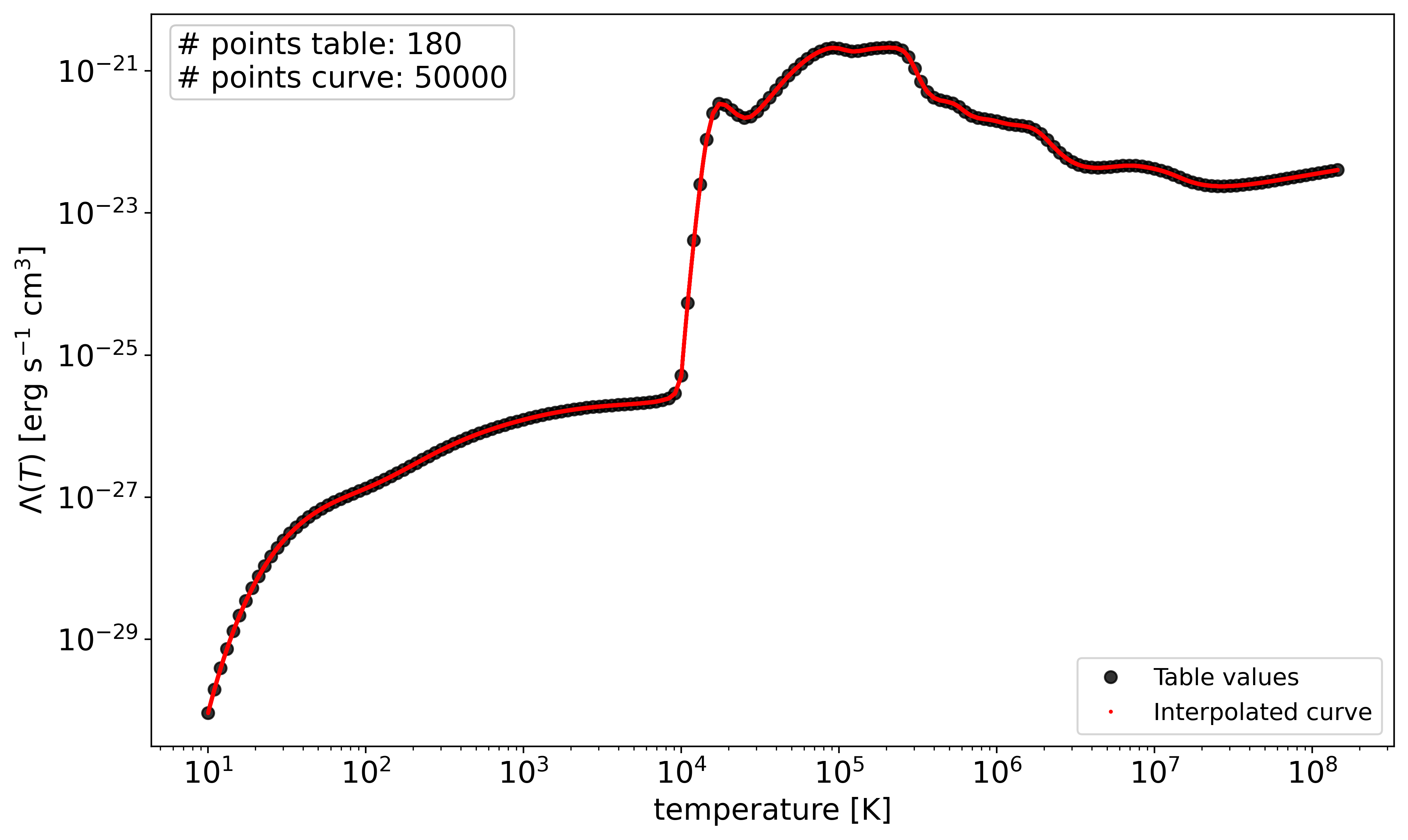}
	\caption{
		Cooling curve $\Lambda(T)$ used in this work, from \citet{schure2009}, extended to low temperatures using \citet{dalgarno1972}. Black dots represent the tabulated values, the red line is a local second-order polynomial approximation at 50000 points.
	}
	\label{fig: coolingcurve}
\end{figure}

We consider the non-adiabatic (dimensionless) MHD equations with the addition of an external gravitational field,
\begin{gather}
	\frac{\partial \rho}{\partial t} = -\nabla \cdot (\rho\bfv),	\label{eq: continuity} \\
	\rho\frac{\partial \bfv}{\partial t} = -\nabla p - \rho\bfv \cdot \nabla \bfv  + (\nabla \times \bfb) \times \bfb + \rho\bfg,	\label{eq: momentum} \\
	\rho\frac{\partial T}{\partial t} = -\rho\bfv \cdot \nabla T - (\gamma - 1)p\nabla\cdot\bfv  - (\gamma - 1)\rho\HL + (\gamma - 1)\nabla\cdot(\bfkappa \cdot \nabla T), \label{eq: energy} \\
	\frac{\partial \bfb}{\partial t} = \nabla \times (\bfv \times \bfb).	\label{eq: induction}
\end{gather}
Here $\rho$, $\bfv$, $T$ and $\bfb$ represent plasma density, velocity, temperature and magnetic field, respectively, with $\bfg$ the gravitational acceleration. $\gamma$ denotes the ratio of specific heats, taken to be $5/3$. Heating and cooling effects through optically thin radiative losses are represented by the heat-loss function $\HL$ \citep{parker1953}, given by
\begin{equation}	\label{eq: heat-loss}
	\HL = \rho\Lambda(T) - \mathcal{H},
\end{equation}
with $\Lambda(T)$ a tabulated set of values resulting from detailed molecular calculations, hereafter referred to as the cooling curve. In this work we will use a cooling curve given by \citet{schure2009}, which is extended to low temperatures using \citet{dalgarno1972} and interpolated at high resolutions using a local second-order polynomial approximation. Both the table values and the interpolated cooling curve are shown in Figure \ref{fig: coolingcurve}. In principle one can take any cooling curve here, but as long as relatively modern cooling tables are used the results will be quite similar \citep{hermans2021}. The symbol $\mathcal{H}$ represents the energy gains and is taken equal to the radiative losses of the equilibrium state, such that the system is ensured to be in thermal equilibrium. It should be noted that $\mathcal{H}$ is assumed to only depend on the equilibrium quantities and may therefore spatially vary over the domain, but has no influence on the density and temperature derivatives of the heat-loss function in Eq. \eqref{eq: heat-loss}.

For thermal conduction we use a full tensor representation, due to the high anisotropy of this effect in magnetized plasmas.
The thermal conductivity tensor $\bfkappa$ is hence given by
\begin{equation}	\label{eq: kappa_tensor}
	\bfkappa = \kappa_\parallel\bfeb\bfeb + \kappa_\bot(\boldsymbol{I} - \bfeb\bfeb),
\end{equation}
in which $\kappa_\parallel$ and $\kappa_\bot$ represent the thermal conduction coefficients parallel and perpendicular to the magnetic field lines, $\boldsymbol{I}$ is the identity matrix and $\bfeb = \bfb / B$ denotes a unit vector along the magnetic field.
For the conduction coefficients themselves we use the Spitzer conductivity as in \citet{legolas_paper_2020}.
Equations \eqref{eq: continuity}-\eqref{eq: induction} are linearized around a static background (without flow) with a one-dimensional variation in the $x$-direction along gravity, meaning the adopted equilibrium state is of the form $\rho_0(x)$, $p_0(x)$, $T_0(x)$. The magnetic field profile is given by
\begin{equation}	\label{eq: equilibrium}
	\begin{aligned}
		\bfb_0(x) &= B_{0y}(x)\bey + B_{0z}(x)\bez,
	\end{aligned}
\end{equation}
that is, the field is horizontal everywhere in our vertically stratified atmosphere, but can have shear and can vary in magnitude.
This naturally implies that $\nabla \cdot \bfb_0 = 0$, hence the background magnetic field is divergence-free. Linearizing around this static equilibrium state and assuming an external gravity field $\bfg = -g(x)\bex$ yields the following system of linearized equations where a subscript 1 denotes the perturbed quantities.
\begin{gather}
	\frac{\partial \rho_1}{\partial t} = -\rho_0 \nabla \cdot \bfv_1,			\label{eq: linearised_continuity}	\\
	\begin{gathered}
		\rho_0 \frac{\partial \bfv_1}{\partial t} = - \nabla\left(\rho_1T_0 + \rho_0T_1\right)
		+ (\nabla \times \bfb_0) \times (\nabla \times \bfA_1)
		+ \left[\nabla \times (\nabla \times \bfA_1)\right] \times \bfb_0 + \rho_1 \bfg,
	\end{gathered}	\label{eq: linearised_momentum}	\\
	\begin{gathered}
		\rho_0\frac{\partial T_1}{\partial t} = -\rho_0\bfv_1 \cdot \nabla T_0
		- (\gamma - 1)\rho_0T_0\nabla \cdot \bfv_1 - (\gamma - 1)\rho_1\HL_0
		- (\gamma - 1)\rho_0\left(\HL_T T_1 + \HL_\rho \rho_1\right)	\\
		+ (\gamma - 1)\nabla \cdot (\bfkappa_0 \cdot \nabla T_1)
		+ (\gamma - 1)\nabla \cdot (\bfkappa_1 \cdot \nabla T_0),
	\end{gathered} 	\label{eq: linearised_energy} \\
	\frac{\partial \bfA_1}{\partial t} = \bfv_1 \times \bfb_0,	\label{eq: linearised_induction}
\end{gather}
Here we adopted a vector potential to write the perturbed magnetic field as $\bfb_1 = \nabla \times \bfA_1$, which in turn ensures that $\nabla \cdot \bfb_1 = 0$. The perturbed pressure $p_1$ was replaced by $\rho_1 T_0 + \rho_0 T_1$ using the linearized ideal gas law, for a detailed treatment on the linearization of the thermal conductivity tensor $\bfkappa$ and the heat-loss function $\HL$ we refer to \citet{legolas_paper_2020}. The quantities $\HL_\rho$ and $\HL_T$ represent the density and temperature partial derivatives of the heat-loss function, respectively.

\subsection{Solving the equations}
A Fourier analysis is done on the 3D, time-dependent perturbed quantities and standard Fourier modes are imposed on the ignorable coordinates $y$ and $z$ with an exponential time dependence, that is,
\begin{equation}	\label{eq: fourier}
	f_1 = \hat{f_1}(x)\exp\left[i(k_y y + k_z z - \omega t)\right],
\end{equation}
where $\omega = \omega_R + i\omega_I$ denotes the (complex) frequency of the wave modes and $f_1$ represents any of the perturbed quantities.
This Fourier decomposition converts the partial differential equations \eqref{eq: linearised_continuity}-\eqref{eq: linearised_induction} into a set of ordinary linear differential equations.
These are solved using the open-source and fully documented \texttt{Legolas} code \citep{legolas_paper_2020, legolas_erratum}, which uses a finite element representation and a weak Galerkin formalism where the unknown functions are approximated by a linear combination of quadratic and Hermite basis functions on a spatially discretized grid. This approach transforms the set of ordinary linear differential equations into a generalized non-Hermitian complex eigenvalue problem of the form $\amat\bfX = \omega\bmat\bfX$, where $\bfX$ denotes the state vector containing the 8 perturbed quantities. In this representation the eigenvalues are given by $\omega$, and due to the assumed exponential time dependence in Eq. \eqref{eq: fourier} this implies that if $\omega_I$ is nonzero the wave modes (with $\omega_R \neq 0$) are either damped (negative $\omega_I$) or overstable (positive $\omega_I$). In the case of $\omega_R = 0$, we similarly have purely damped or purely unstable modes.

\texttt{Legolas} can solve the aforementioned eigenvalue problem using a QR-algorithm based on LAPACK routines \citep{lapack}, which yields the full spectrum (that is, all eigenvalues at that resolution) and corresponding eigenfunctions. A second possibility is solving through use of ARPACK, which relies on an implicitly restarted Arnoldi iteration \citep{arpack} to calculate a user-set number of eigenvalues with specific features (usually the ones with largest magnitude but other options are possible as well). The latter can either be done on the entire spectrum itself, or around a given point in the complex plane through use of a shift-invert method which yields those eigenvalues in the immediate vicinity of a chosen reference (complex) target frequency.

\subsection{Non-adiabatic continua}
A well-known feature of linear ideal MHD is the existence of slow and Alfv\'en continua, corresponding to singular points in the equivalent formation of the linear partial differential equations from Eqs. \eqref{eq: continuity}-\eqref{eq: induction} into an ordinary differential equation in terms of the $x$-component of the Lagrangian displacement field. In essence, these continua represent finite ranges of eigenfrequencies with singular, ultra-localized eigenfunctions, that play a key role in processes like phase mixing, resonant absorption, or uniturbulence processes. In ideal MHD, the forward- and backward slow and Alfv\'en continua are always real (pure waves), and the eigenfrequency ranges are $x$-dependent in our formalism. Viewed as a function of $x$, local minima or maxima in the continua may lead to the presence of additional, discrete sequences of modes that cluster to these accumulation points. For a detailed discussion on this topic we refer to \citet{book_MHD}. The inclusion of non-adiabatic effects introduces the thermal continuum, which in static, non-adiabatic settings is just the marginal entropy mode. It was shown by \citet{vanderlinden1991} that these continuum modes are introduced by $\kappa_\bot = 0$, and hence only exist if there is no perpendicular thermal conduction. For nonzero $\kappa_\bot$ the thermal continuum gets replaced by a quasi-continuum, represented by a dense band of thermal modes.
Furthermore isolated instabilities can also be found, as well as accumulation sequences that can be predicted by WKB analysis, also for the non-adiabatic regime \citep{keppens1993}.

The Alfv\'en continuum remains unmodified when only non-adiabatic effects are included and hence all its continuum solutions will be real, corresponding to locally resonant waves. The slow continuum on the other hand is modified and couples to the thermal continuum, with solutions given by a third order polynomial in $\omega$ as derived in \citet{vanderlinden1991} and adapted to our current formalism,
\begin{equation}	\label{eq: continuum_modes}
	\begin{gathered}
		\frac{\rho_0 \left(c_s^2 + c_A^2\right)}{\gamma - 1}i\omega^3
		+ \left[\Qaifull\right]\omega^2
		- \frac{\rho_0 c_s^2c_A^2 k_\parallel^2}{\gamma - 1}i\omega \\
		- \left[\Qifull\right]c_A^2 k_\parallel^2  = 0,
	\end{gathered}
\end{equation}
where $c_s^2 = \gamma p_0 / \rho_0$ and $c_i^2 = p_0 / \rho_0$ represent the sound speed and isothermal sound speed, respectively, and $c_A^2 = B_0^2 / \rho_0$ denotes the Alfv\'en speed. The quantity $k_\parallel = \bfk \cdot \bfb_0 / B_0$ represents the wave vector component parallel to the magnetic field, everything is normalized. In general, Equation \eqref{eq: continuum_modes} usually has one purely imaginary solution, corresponding to the thermal continuum $\omega_{th}$, and two complex conjugate solutions representing the slow continua $\omega_S^\pm$. In the ideal case $\omega_{th}$ vanishes (reducing to a marginal entropy solution) while the slow continua reduce to their ideal counterparts $\omega_S^2 = k_\parallel^2 c_s^2 c_A^2 / (c_s^2 + c_A^2)$, which may in turn collapse into points if $c_s$ and $c_A$ are both not spatially varying.
If the slow continuum vanishes, which is the case if $\bfk \cdot \bfb_0 = 0$ or if a pressureless background is considered (that is, adopting a zero plasma-beta), the thermal continuum has an analytical solution given by
\begin{equation}
	\omega_{th} = \frac{(\gamma - 1)i}{c_s^2 + c_A^2}\left[\rho_0\HL_\rho - \HL_T\left(c_i^2 + c_A^2\right)\right],
\end{equation}
and hence becomes independent of thermal conduction effects.

It is important to stress that continuum modes are in fact actual solutions to the eigenvalue problem, and hence represent physical modes. Consequently, one can draw the following general conclusions for the thermal stability of a given equilibrium, solely based on the behavior of its slow and thermal continuum regions:
\begin{enumerate}
	\item The continuum regions are partially or fully unstable, that is, at least some modes have a positive imaginary part. Since continuum modes are physical solutions, this means that the medium will be susceptible to instability, either through an unstable slow or thermal mode (or a coalescence of both) depending on which continuum region is unstable. How fast possible resulting condensations form will (usually) depend on the growth rates of the most unstable mode(s), although mode interactions and nonlinear effects can develop quickly.
	\item The continuum regions have no internal extrema and are completely stable, that is, all modes have a negative or vanishing imaginary part. This corresponds to a thermally stable case, without the possibility of forming condensations through thermal instability.
	\item The continuum regions have internal extrema. This represents the more general case, in which discrete modes in the vicinity of the continua \textit{may} exist. Here it is possible to have a thermally stable continuum with an unstable discrete mode such that the medium is still thermally unstable.
\end{enumerate}
For the solar atmosphere we are usually dealing with case three, since the strong temperature and density variations invariably lead to sharp gradients in both the heat-loss function and its derivatives, resulting in continuum regions with internal extrema.
In the remainder of this paper, we will show that realistic models indeed show a great variety of linear pathways to forming condensations. That the coupled thermal and slow continua, with their ultra-localized eigenfunctions, are true solutions that will play a role in solving initial value problems, is thus a clear challenge to any nonlinear simulation model.

\section{Comparison with analytical work} \label{sec: analytical}
Before going into fully realistic solar atmosphere setups, we first revisit a solar atmospheric model described by \citet{nye1976}. These authors considered an isothermal atmosphere with a horizontal magnetic field independent of height and an exponentially decreasing density profile in a uniform gravitational field. This particular setup results in a constant sound speed and density scale height, which, combined with a wave vector parallel to the magnetic field, makes it analytically tractable. The equilibrium background (in normalized units) is given by
\begin{equation} \label{eq: nye-thomas}
	\begin{gathered}
		c_s^2 = \gamma T_0, \qquad
		H = \frac{T_0}{g} = \text{const.}, \qquad
		\rho_0(x) = \rho_{00}e^{-x/H}, \qquad
		B_z = \sqrt{\gamma \rho_{00}T_0 \beta^2} = \text{const.},
	\end{gathered}
\end{equation}
where $H$ denotes the density scale height, $\rho_{00}$ is taken equal to one and $\beta^2 = c_{A0}^2 / c_s^2$ is a dimensionless parameter defined as the ratio between the Alfvén velocity at $x = 0$ and the (constant) sound speed. Furthermore solar gravity is assumed, hence the dimensional value for $g = 274$ ms$^{-2}$ ($\approx 1.66$ in normalized units). In our setup we take a reference scale height of one and use that to constrain the background temperature. \citet{nye1976} then proceeded in deriving a dispersion relation for waves with a wavenumber $k_z$ purely aligned with the constant horizontal magnetic field, which was evaluated for various values of $Hk_z$ and $\beta^2$. We can plug in the equilibrium configuration defined in Eq. \eqref{eq: nye-thomas} in \texttt{Legolas} which will return the full eigenvalue spectrum with all eigenfunctions. It should be noted that the original work only considers modes in the fast sequence, while in our full spectrum calculation also the slow and Alfvén sequences are present.
Also, the analytic work assumes an infinite atmosphere with exponential vanishing eigenfunctions bounded by a solid wall at the lower boundary, while here we have a finite slab with solid wall boundaries on both sides. We therefore take the slabsize sufficiently large (about 15 scale heights) to approximate an infinite atmosphere. All runs in this section are performed using a resolution of 250 gridpoints and unit normalizations taken as 1 MK, 10 Gauss and 50 Mm for the unit temperature, unit magnetic field and unit length, respectively. We first study ideal MHD spectra in Sections \ref{ss: nye-ky0} and \ref{ss: nye-kynonzero}, to then extend the results to non-adiabatic MHD in Section \ref{ss: nye-na}.

\subsection{Revisiting $k_y = 0$} \label{ss: nye-ky0}
\begin{figure}
	\centering
	\includegraphics[width=\textwidth]{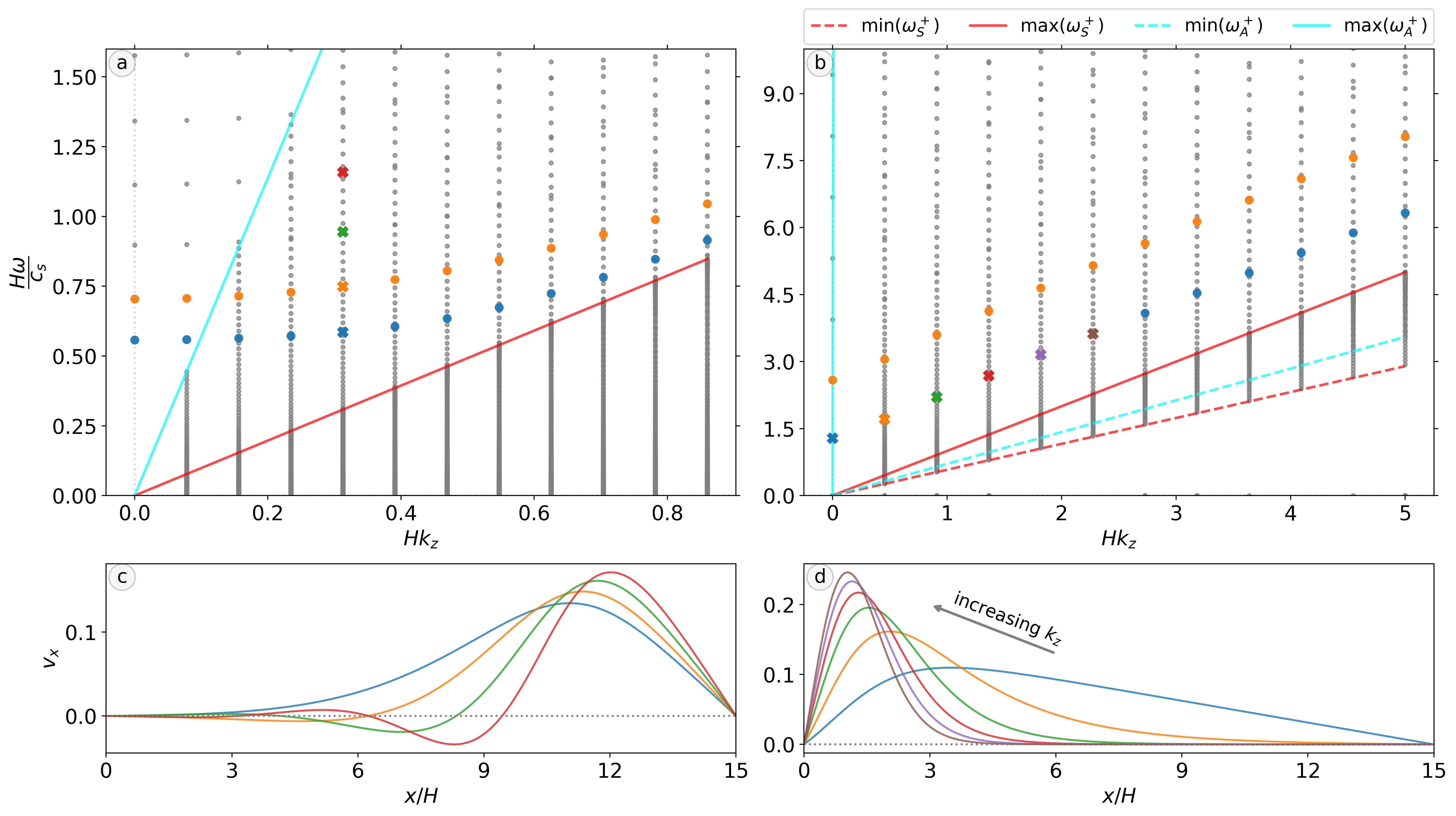}
	\caption{Spectrum plots with eigenfunctions for the equilibrium in Eq. \eqref{eq: nye-thomas}, for $\beta^2 = 10^{-4}$ (panel $a$) and $\beta^2 = 0.5$ (panel $b$), $k_y = 0$ in both cases.
		The blue and orange dots denote the first and second modes of the fast sequence, respectively.
		The solid (dashed) red and cyan lines indicate the maximum (minimum) of the slow and Alfvén continua, respectively. Panel $c$: $v_x$ eigenfunctions of the 4 modes annotated with a cross on panel $a$, near $Hk_z \approx 0.3$,
		representing the first four modes of that fast sequence. Panel $d$: $v_x$ eigenfunctions of the modes marked with crosses on panel $b$, becoming more localized as $k_z$ increases. All eigenfunctions are normalized, the colors are consistent
		between the panels.}
	\label{fig: nye-thomas-ky0}
\end{figure}
Figure \ref{fig: nye-thomas-ky0} shows spectrum plots for $\beta^2 = 10^{-4}$ (panel $a$) and $\beta^2 = 0.5$ (panel $b$), using the equilibrium given in Eq. \eqref{eq: nye-thomas}. The full spectrum is shown with dots, where every vertical line of dots represents one \texttt{Legolas} run for that particular $k_z$ value. The first two modes of the fast sequence are annotated in blue and orange, respectively. The solid red and cyan lines denote the maximum of the slow and Alfvén continuum, respectively, with their minima near zero (but finite). Hence, everything to the left of the cyan line in panel $a$ are purely fast modes, the region between the cyan and red line contains fast modes overlapping with the Alfvén continuum, and the dense collection of points below the red line corresponds to an overlapping between the slow and Alfvén continua. Panel $c$ shows the $v_x$ eigenfunctions of the first four modes in the fast sequence at $Hk_z \approx 0.3$ (that is, the crosses in blue, orange, green and red, respectively, at that particular wave number). Since the magnetic field scales with the parameter $\beta$ (see Eq. \eqref{eq: nye-thomas}), this case is approximately hydrodynamic.

Panel $b$ in Figure \ref{fig: nye-thomas-ky0} has a dynamically important magnetic field ($\beta^2 = 0.5$) and shows a similar setup as panel $a$. The slow and Alfvén continua have their minima annotated with red and cyan dashed lines, respectively. The slow continuum is constrained to a narrow band, with the Alfvén continuum partially overlapping. The maximum value of the latter lies far beyond the upper limit of that panel. Hence, only the modes at $k_z = 0$ represent purely fast modes, all others above the solid red line are overlapping with the Alfvén continuum. The first two modes of each fast sequence are again annotated in blue and orange, respectively. The $v_x$ eigenfunctions in panel $d$ correspond to the six annotated modes on panel $b$, and become more and more localized when $k_z$ increases.

It should be noted that the first sequence of modes in Figure \ref{fig: nye-thomas-ky0} does in fact correspond to the $k_y = 0$ and $k_z = 0$ case. This implies that the slow and Alfv\'en continua collapse to $\omega = 0$, and one could mistakenly be under the impression that there would be no modes present for this case. However, the assumed Fourier dependence in Eq. \eqref{eq: fourier} does in fact allow for nontrivial $\omega$ solutions if $k_y = k_z = 0$ and $\hat{f_1}(x) \neq 0$, representing fast modes propagating in the $x$-direction perpendicular to the magnetic field.

Both spectra show an excellent correspondence to the plots given in \citet{nye1976}, which were obtained by analytically solving the dispersion relation. However, \citet{nye1976} never mentioned the presence of the slow and Alfv\'en continua, which were not described by their dispersion relation.

\subsection{Effects of a nonzero $k_y$} \label{ss: nye-kynonzero}
\begin{figure}[t]
	\centering
	\includegraphics[width=0.6\columnwidth]{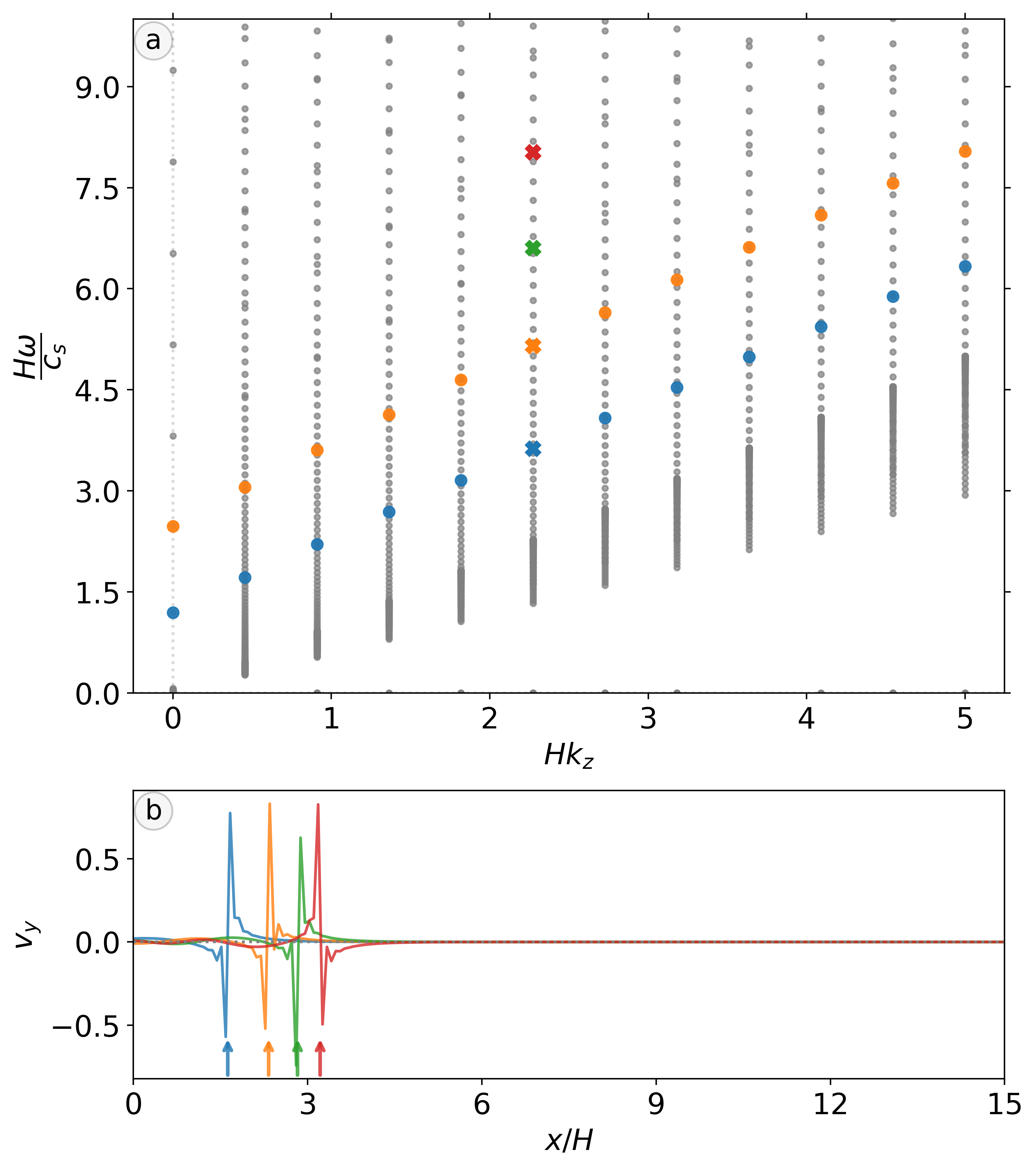}
	\caption{Panel $a$: spectrum plot similar to panel $b$ of Figure \ref{fig: nye-thomas-ky0}, but with $k_y = 0.05$. Blue and orange dots denote the first and second mode of the fast sequences, respectively.
		Panel $b$ shows $v_y$ eigenfunctions of the four modes marked with crosses in the fast sequence at $Hk_z \approx 2.2$, clearly showing the $1/x$ behavior at $x = x_A$ due to overlap and resonance with the Alfvén continuum.}
	\label{fig: nye-thomas-kynonzero}
\end{figure}
As seen in Figure \ref{fig: nye-thomas-ky0} there is an overlap between the fast mode sequence and the Alfvén continuum. However, when the wave vector is taken to be parallel to the magnetic field, the Alfvén continuum is no longer relevant in the sense that it completely decouples from the fast modes. This is the reason why that particular case becomes analytically tractable, even though the background equilibrium is inhomogeneous. However, there is no need to force that condition when using \texttt{Legolas}, and hence we can introduce a nonzero value for $k_y$. Even a small value of $k_y$ will have considerable effects due to the overlap with the Alfvén continuum, since the latter will no longer decouple, resulting in a resonance between the fast modes and the continuum.
It can be shown using a Frobenius expansion (for an excellent discussion on this topic we refer to \citet{book_MHD}) that this will introduce a singularity with a $(x - x_A)^{-1}$ behavior in the eigenfunctions perpendicular to the magnetic field, where $x_A$ is the resonance radius. In the case considered here it is actually possible to make an estimate of this radius, by starting from the regular expression for the Alfvén continuum $\omega_A^+(x) = (\bfk \cdot \bfb_0) / \sqrt{\rho(x)}$. Since the magnetic field is homogeneous, this expression can be inverted to yield $x_A$ as a function of frequency:
\begin{equation}	\label{eq: xa_estimate}
	x_A = 2 H \ln\left(\frac{\omega}{k_z B_z}\right),
\end{equation}
where we used the fact that $\rho_{00} = 1$ and the magnetic field is aligned with the $z$-axis. If we now substitute the frequency of the fast mode under consideration, expression \eqref{eq: xa_estimate} will give an indication where in the eigenfunction the $1/x$ singularity will occur. Naturally, this also implies that the location of this singularity is frequency and wave number dependent, resulting in a larger $x_A$ when the frequency increases (that is, for higher modes in the sequence) and a smaller $x_A$ when $k_z$ increases.

Figure \ref{fig: nye-thomas-kynonzero} shows the same setup as panel $b$ in Figure \ref{fig: nye-thomas-ky0}, but now with $k_y = 0.05$. As seen on panel $a$ the spectrum itself is practically unaffected, the first two modes of each fast sequence are again annotated.
Panel $b$ shows the $v_y$ eigenfunctions for the four modes at $Hk_z \approx 2.2$, clearly showing the $1/x$ singularity. The arrows denote the estimated value for $x_A$ according to Eq. \eqref{eq: xa_estimate} in the same color as its eigenfunction, with values given by $x_A \approx 1.63$, $x_A \approx 2.33$, $x_A \approx 2.83$ and $x_A \approx 3.22$ for the first, second, third and fourth mode of the fast sequence at that value for $k_z$, respectively. Since these modes correspond to the same fast mode sequence, and thus have the same value for $k_z$, we indeed see the singularity moving to the right, locating itself at larger $x_A$ values for higher mode numbers.

It should be noted that these eigenfunctions do not show a ``real" $1/x$ behavior in the sense that the eigenfunctions rather show a spiked oscillation instead of extending towards infinity. This is due to \texttt{Legolas} trying to resolve those eigenfunctions with the limited number of gridpoints at its disposal. The more the resolution is increased, the more this singular $1/x$ profile will be resolved. This truly singular behavior due to this resonance is particular to ideal MHD, but we could also activate resistive terms in \texttt{Legolas}, to then get a fully resolved complex eigenfrequency at finite resistivity with a spatially resolved eigenfunction behavior.

\begin{figure}[b!]
	\centering
	\includegraphics[width=\textwidth]{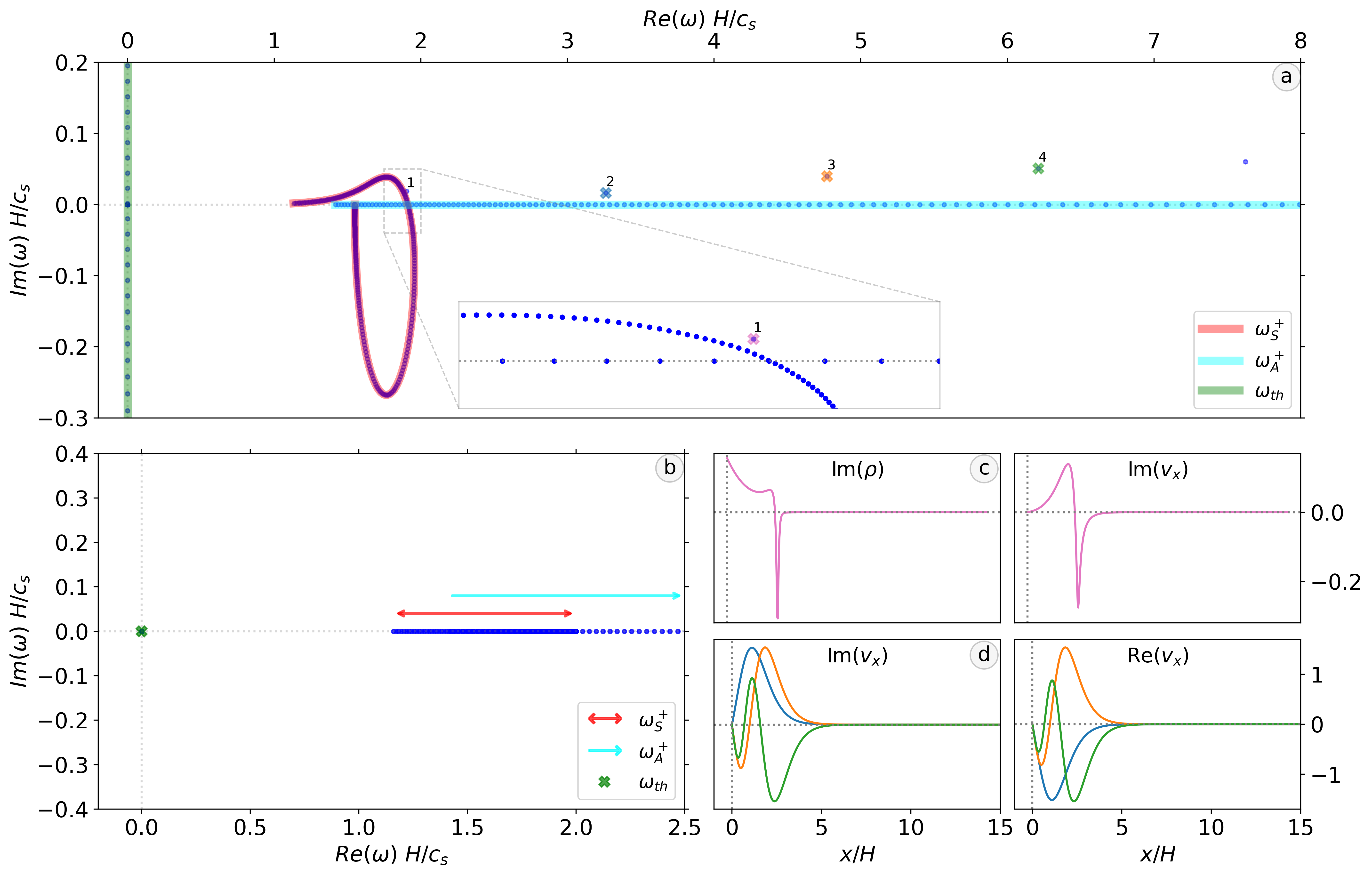}
	\caption{Spectra and eigenfunctions for the $\beta^2 = 0.5$ setup in Section \ref{ss: nye-ky0}, with non-adiabatic effects included.
		Panel $a$: zoom-in near the slow sequence, showing (part of) the thermal, slow and Alfv\'en continua in green, red and cyan, respectively, along with four unstable modes from the fast sequence.
		Inset: further zoom on the slow continuum, revealing a single discrete mode.
		Panel $b$: same as in panel $a$, but without non-adiabatic effects. Panels $c$: $\rho$ (left) and $v_x$ (right) eigenfunctions of the discrete mode annotated in the inset.
		Panels $d$: $v_x$ eigenfunctions for the three fast modes annotated on panel $a$. All eigenfunctions are normalized.}
	\label{fig: nye-thomas-NA}
\end{figure}

\subsection{Inclusion of non-adiabatic effects} \label{ss: nye-na}
\begin{figure}[t]
	\centering
	\includegraphics[width=0.5\columnwidth]{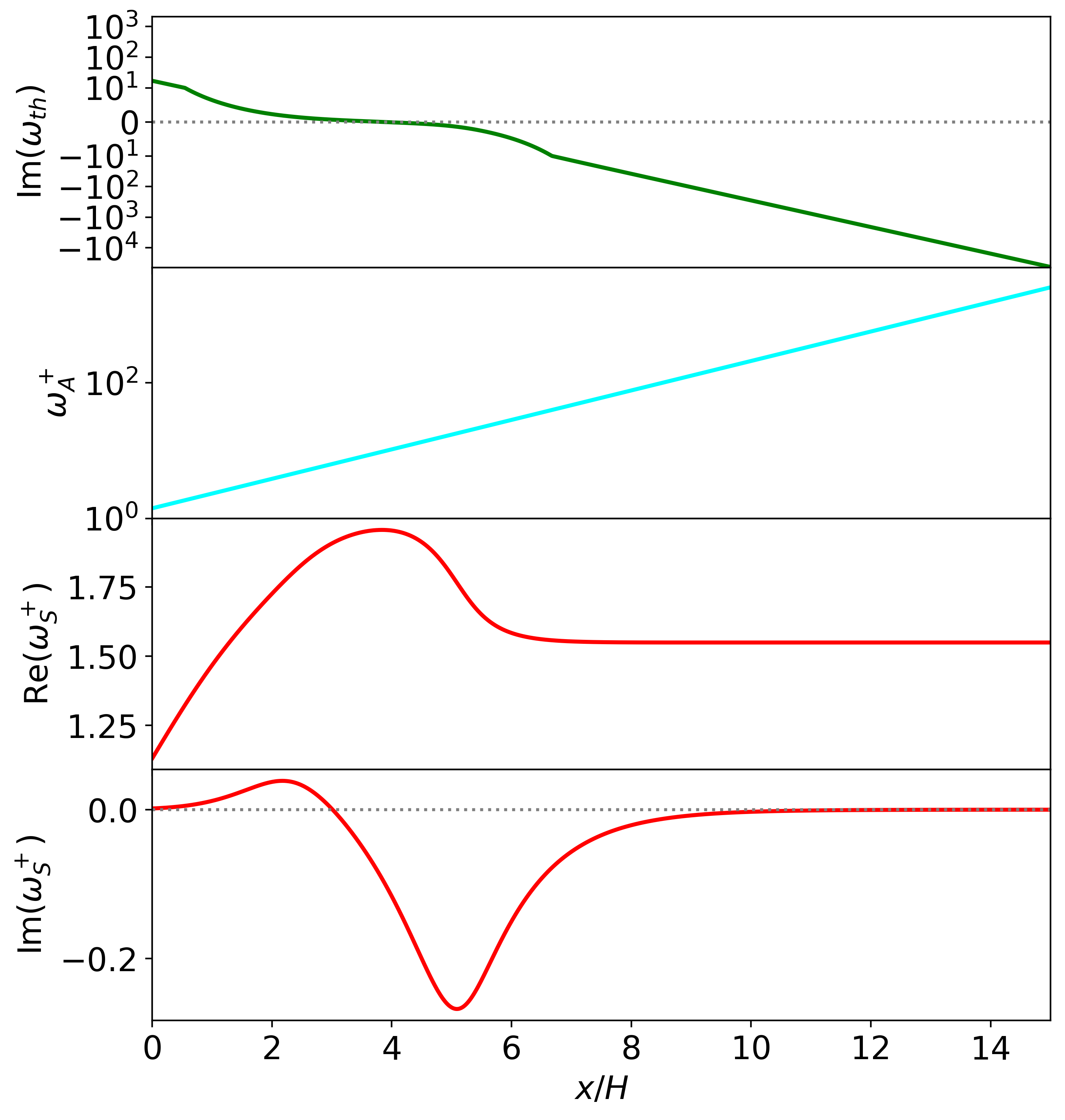}
	\caption{The thermal (green), Alfv\'en (cyan) and slow (red) continua as a function of the grid, for the non-adiabatic case shown in Figure \ref{fig: nye-thomas-NA}.
			 The slow continuum shows an internal extremum. All continua are rescaled to $H / c_s$, similar to the other figures in this Section.}
	\label{fig: nye-thomas-continua}
\end{figure}

As a final extension to this rather simple model we again consider the $\beta^2 = 0.5$ configuration from Section \ref{ss: nye-ky0} but now include non-adiabatic effects: thermal conduction parallel to the magnetic field lines and optically thin radiative losses based
on the cooling curve given in Figure \ref{fig: coolingcurve}. Since this will cause the introduction of complex eigenfrequencies, we now focus on the full spectrum for $k_z = 2$ and $k_y = 0$. Figure \ref{fig: nye-thomas-NA} shows the corresponding spectrum for 250 grid points in panel $a$, whereas panel $b$ shows the exact same spectrum but without non-adiabatic effects for comparison. The thermal, slow and Alfv\'en continua are annotated in green, red and cyan, respectively. By comparing panels $a$ and $b$ it can be seen that the thermal continuum shifts from its marginal solution in the adiabatic case to the dense band of modes on the imaginary axis following Eq. \eqref{eq: continuum_modes}, in this case with both a stable and unstable part. The same holds true for the slow continuum, which shifts from its purely real adiabatic solution to an intricate loop in the complex plane, becoming partially unstable. The inset on panel $a$ zooms in on the slow sequence, revealing a single discrete mode annotated by a pink cross, with the imaginary part of its $\rho$ and $v_x$ eigenfunctions shown in pink on panels $c$, both showing a strong oscillation in the lower regions of the slab. This discrete eigenvalue represents a global unstable slow mode since it is not part of the continuum, although it can be linked to an internal extremum (see Figure \ref{fig: nye-thomas-continua}).

The first few (unstable) modes of the fast sequence in panel $a$ overlap with the Alfv\'en continuum, which is decoupled due to $k_y$ being zero. The real part of these fast modes is approximately equal to the values shown on Figure \ref{fig: nye-thomas-ky0}, panel $b$ (the first and second colored modes on that panel would correspond to modes $2$ and $3$ on Figure \ref{fig: nye-thomas-NA}, panel $a$); the imaginary part is now non-zero due to the non-adiabatic effects included. The $v_x$ eigenfunctions of these fast modes are given in panels $d$. All eigenfunctions on this figure are calculated using a shift-invert method near the eigenvalues of interest.

Figure \ref{fig: nye-thomas-continua} shows all continua corresponding to the non-adiabatic case discussed in this Section, rescaled using the scale height and sound speed in order to be consistent with Figure \ref{fig: nye-thomas-NA}.
The thermal continuum (green) has clear variation, but has no internal extrema. The slow continuum on the other hand does have an internal maximum in its real part, which can be linked back to the unstable discrete slow mode we found, shown in the inset on panel $a$, Fig. \ref{fig: nye-thomas-NA}. Note that this simple magnetized atmosphere model thus already gives us unstable parts in the thermal and slow continuum, and gives overstable discrete modes linked to both the slow and the fast mode sequences.

What this implies for an initial value problem where this particular atmosphere would be realized in a numerical nonlinear MHD simulation with all these non-adiabatic effects included is as follows.
Since the thermal continuum has an unstable part which is almost two orders of magnitude larger than the unstable parts of either the slow or fast modes, one would expect that the most unstable thermal modes would dominate the thermal behavior since these have the largest growth rates. Physically, this would mean that any perturbation would immediately lead to an in-situ condensation. However, the possibility of some form of coalescence between the unstable slow and fast modes and the thermal modes (or a combination of all three) can not be excluded. In that case a cool and dense condensation will still form, but the way this happens (and its subsequent dynamics) may be much more intricate than a ``simple" in-place condensation.

\section{A realistic solar atmosphere model} \label{sec: solar-atmo}
We will now extend the study of waves and instabilities in stratified, magnetized atmospheres to a fully realistic atmosphere description. This requires us to specify the equilibrium (Section \ref{ss: setup}), after which we first look at the non-adiabatic continua governed by Equation \eqref{eq: continuum_modes}. Then, we show full MHD spectra for the solar atmosphere, focusing on specific regions of interest. All models in this Section assume solid wall boundaries on both sides of the slab.
\subsection{The stratified, magnetized atmosphere} \label{ss: setup}
The choice of equilibrium profiles in Eq. \eqref{eq: equilibrium} in combination with the system of equations considered yields two conditions for the unperturbed parts. These equilibrium conditions are necessary to ensure a force-balanced state and are given by
\begin{align}
	\left(\rho_0 T_0 + \frac{1}{2}\bfb_0^2\right)' &= -\rho_0 g, \label{eq: force-balance} \\
	\left(\kappa_\bot T_0'\right)' &= \rho_0 \HL_0, \label{eq: thermal-balance}
\end{align}
adapted from their general formalism in \citet{legolas_paper_2020} to a Cartesian geometry, the prime corresponds to the derivative with respect to the height coordinate $x$.
We are free to choose any kind of equilibrium prescription, as long as these conditions are fulfilled.
Here $\HL_0$ represents the heat-loss function of the equilibrium state, which is zero due to the fact that we assume the heating function $\mathcal{H}$ in Eq. \eqref{eq: heat-loss} such that it balances out the radiative losses, thereby ensuring thermal balance. Hence, since we ignore the finite perpendicular conduction coefficient $\kappa_\bot$, Eq. \eqref{eq: thermal-balance} is automatically satisfied.

\begin{figure}[t]
	\centering
	\includegraphics[width=0.7\columnwidth]{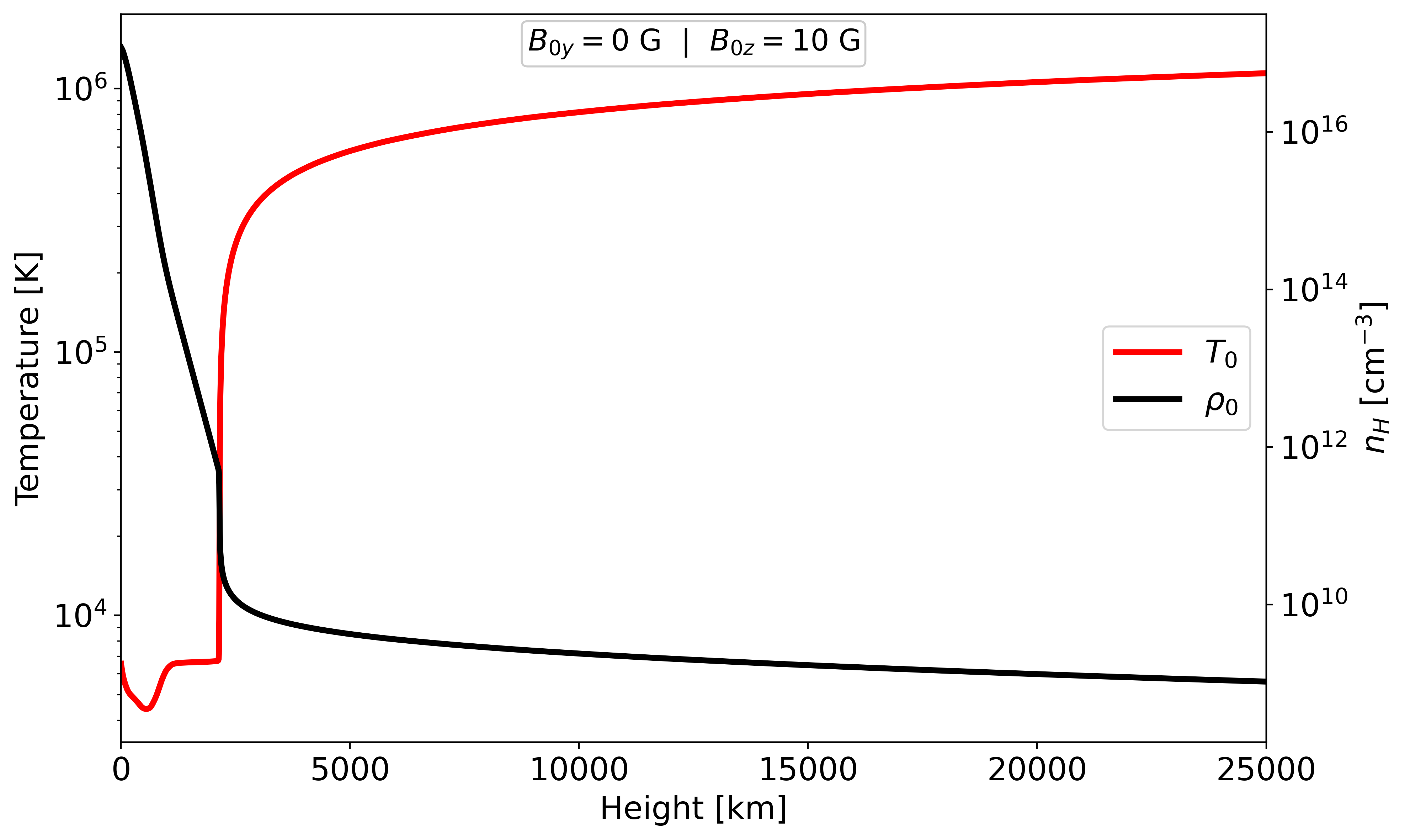}
	\caption{Equilibrium temperature profile (red) based on the solar atmospheric model by \citet{avrett2008}, with corresponding integrated density profile ($n_H$, black) from the ODE in Eq. \eqref{eq: rho_profile}. The magnetic field has a uniform $z$-component of 10 G, the
		gravitational field is given by Eq. \eqref{eq: gfield}.}
	\label{fig: profile}
\end{figure}

In our model considered here, we assume a Cartesian slab parallel to the solar surface infinitely extended along the $y$ and $z$ directions, with a variation along $x$ in the equilibrium quantities and an external gravitational field directed downward towards the solar surface. For the temperature profile we use a solar atmospheric model proposed by \citet{avrett2008}, which is based on detailed radiative transfer calculations and observations. This profile is interpolated at high resolution using a local second-order polynomial approximation, similar to how the radiative cooling curve is treated (see Figure \ref{fig: coolingcurve}). For the gravitational field we assume the following profile,
\begin{equation}	\label{eq: gfield}
	g = g_\odot\left(\frac{R_\odot}{R_\odot + x}\right)^2,
\end{equation}
where $g_\odot = 274$ ms$^{-2}$ denotes the solar gravitational constant and $R_\odot = 6.96\times 10^{10}$ cm the solar radius. The magnetic field is assumed to be uniform and horizontal (aligned along $z$). Note that this is similar to the analytic case discussed before, and still avoids horizontal magnetic fields that shear or vary in magnitude with height. Also, no vertical equilibrium magnetic field components are considered here. Since this kind of equilibrium is numerical, the density profile has to be calculated in order to achieve a force-balanced state. Once the temperature, gravity and magnetic field profiles are chosen, Eq. \eqref{eq: force-balance} can be rewritten in terms of density as the following first order ordinary differential equation
\begin{equation}	\label{eq: rho_profile}
	\rho'(x) = -\frac{T'(x) + g(x)}{T(x)}\rho(x) - \frac{B_{0y}(x)B_{0y}'(x) + B_{0z}(x)B_{0z}'(x)}{T(x)}.
\end{equation}
The uniform magnetic field adopted implies that only the first term here enters, but one could allow for non-uniform, non force-free magnetic field variations by including this second term.
The temperature derivative has to be done numerically and is calculated based on the high-resolution interpolated profile using sixth order accurate finite difference methods, with forward and backward differences near the start and end of the grid, respectively, and central differences in between. Equation \eqref{eq: rho_profile} is then integrated for $\rho(x)$ using a fifth-order accurate Runge-Kutta method to obtain the density profile, which is then substituted back into the equation in order to retrieve the density derivative. This approach in calculating the density derivative (instead of numerically differentiating the obtained density profile) satisfies the force-balance equation \eqref{eq: force-balance} as close as numerically possible.
As unit normalizations we choose a reference temperature of 1 MK, magnetic field of 10 Gauss and a unit length of 1000 km which automatically constrain all other normalizations following the ideal gas law using a mean molecular weight of one.
This results in a profile which corresponds to observational values of the solar atmosphere as given in \citet{book_priest}.

Figure \ref{fig: profile} shows the resulting temperature and density profiles as a function of position, extending up to 25 Mm above the solar surface, for a uniform magnetic field of 10 Gauss along the $z$-direction. These equilibria are subsequently passed on to \texttt{Legolas} with a given wave vector to calculate the spectra. Thermal conduction parallel to the magnetic field is included, together with optically thin radiative losses based on the cooling curve by \citet{schure2009} as in Figure \ref{fig: coolingcurve}.

\subsection{Non-adiabatic slow \& thermal continua}
\begin{figure}[!htb]
	\centering
	\includegraphics[width=\textwidth]{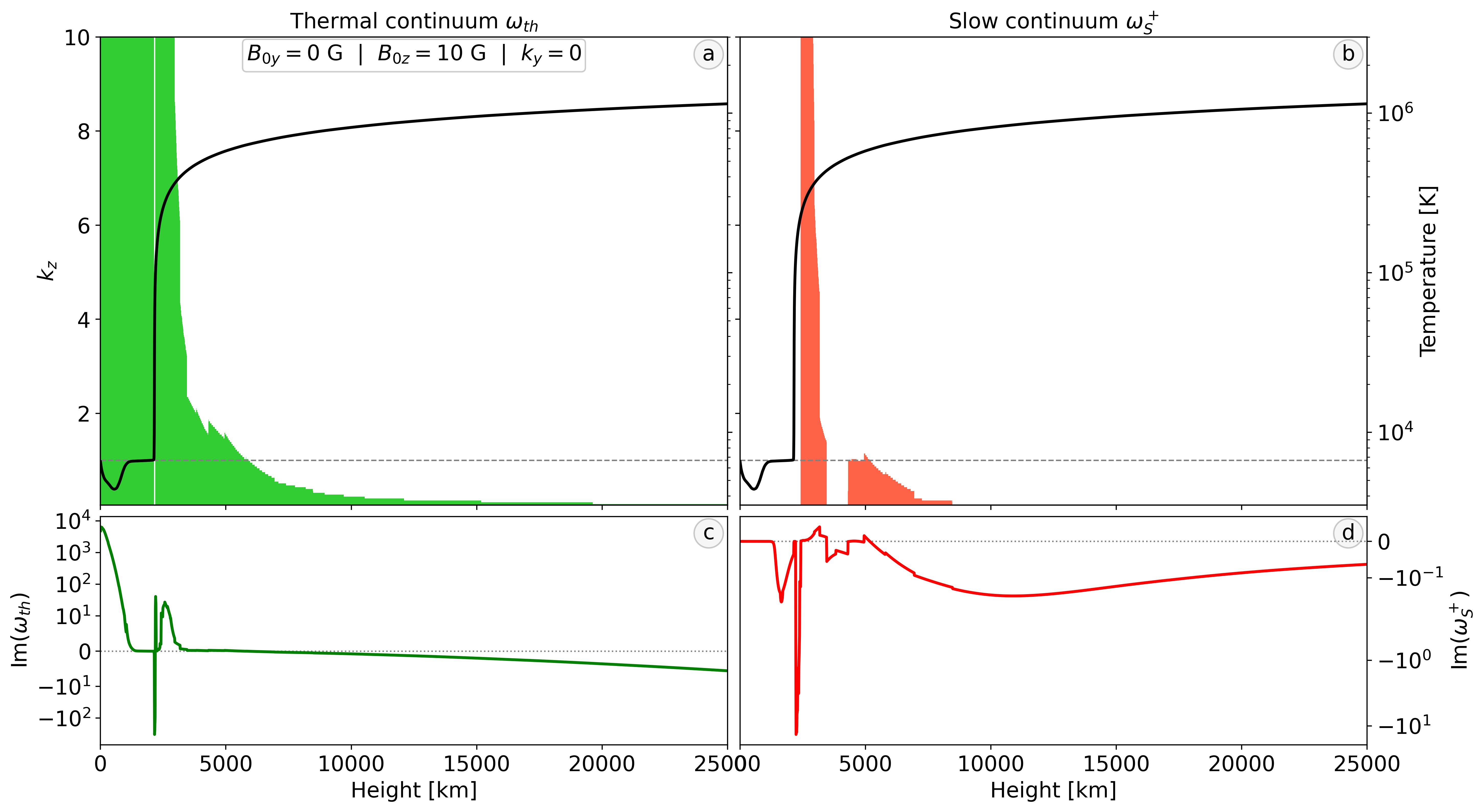}
	\caption{Stability regions of the thermal and slow continua for the temperature profile in Fig. \ref{fig: profile}, for a range in $k_z$. Unstable regions are colored in green and red for the thermal and slow continua, respectively.
		Panels $c$ and $d$: continuum solutions as a function of height for $k_z = 1$, corresponding to the dotted line in panels $a$ and $b$.}
	\label{fig: continua-kzvary}
\end{figure}

Before diving into the complex spectroscopy of a realistic solar atmosphere model, it is useful to have a first indication of stable and unstable configurations in order to locate possible parameter regions of interest.
To that end we solve Equation \eqref{eq: continuum_modes} using the profiles discussed in Section \ref{ss: setup} for a range in wave numbers. Figure \ref{fig: continua-kzvary} shows stability regions for both the thermal and slow continua over the entire temperature range shown in Fig. \ref{fig: profile}, for a wave vector parallel to the magnetic field ($k_y = 0$) with $k_z$ varying between $0.05$ and $10$. The temperature profile is superimposed on both figures with a black line. Regions where the continua are unstable are colored in green for the thermal continuum (panel $a$) and in red for the slow continuum (panel $b$). It can immediately be seen that the entire chromosphere up to the transition region at $\approx 2$ Mm has an unstable thermal continuum for the full range in wave number, and as soon as the lower corona is reached this becomes more and more stable, until only large-scale perturbations (small $k_z$) lead to unstable continuum regions.
The high thermal instability of the solar chromosphere is mainly a consequence of the low temperature present there, which results in diminished parallel thermal conduction effects ($\sim T^{5/2}$) and hence less stability since temperature gradients are less efficiently smoothed out. The slow continuum on the other hand is completely stable in the chromosphere and is characterized by only a narrow unstable region in the lower corona.
Panels $c$ and $d$ correspond to the horizontally dotted line at $k_z = 1$, and show the imaginary parts of the thermal ($c$) and slow ($d$) continua for that particular $k_z$ value as a function of height.

\begin{figure}[!htb]
	\centering
	\includegraphics[width=\textwidth]{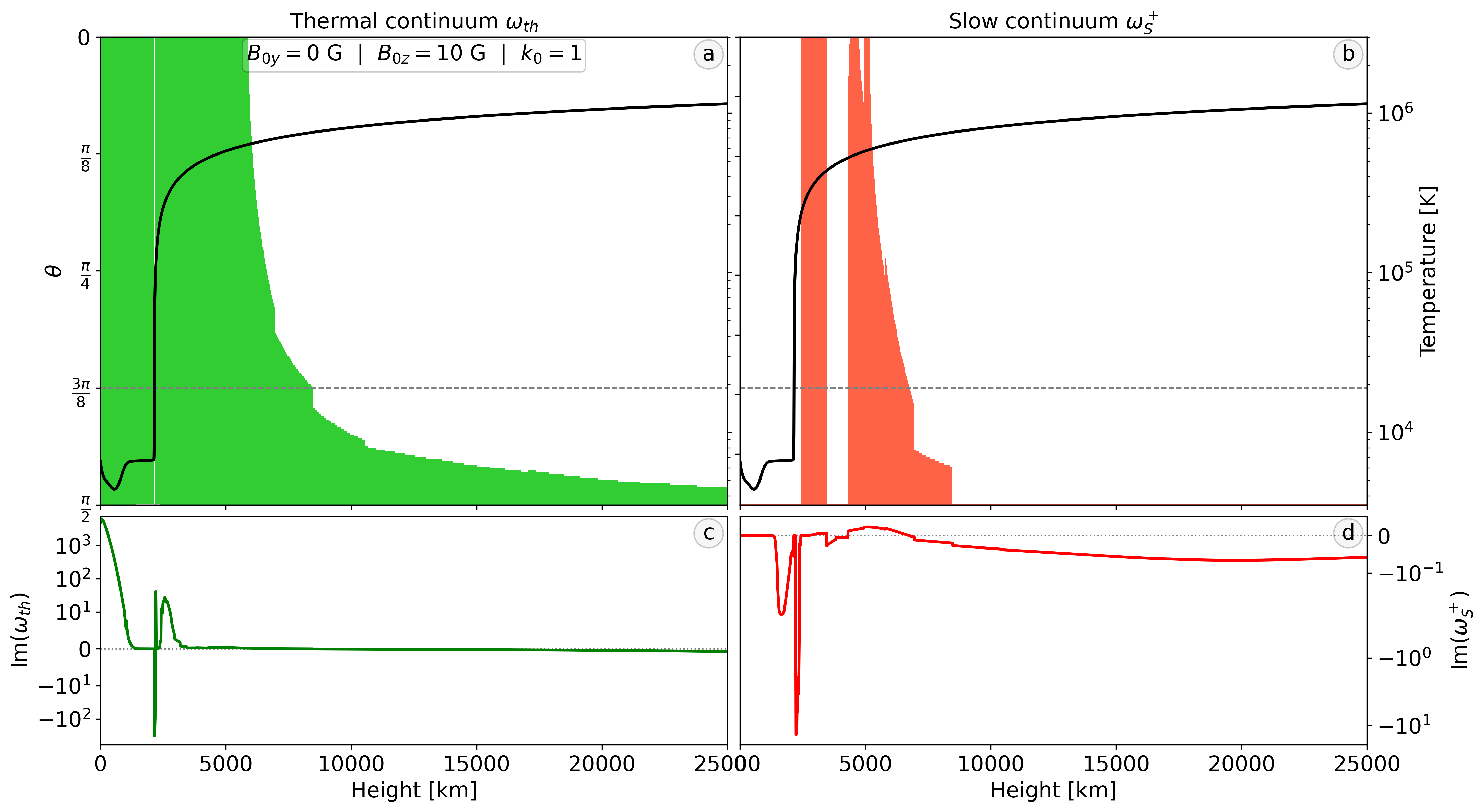}
	\caption{Stability regions of the thermal and slow continua for the temperature profile in Fig. \ref{fig: profile}, for various angles between the wave vector $\bfk$ and the magnetic field $\bfb_0$.
		Unstable regions are colored in green and red for the thermal and slow continua, respectively. Panels $c$ and $d$: continuum solutions as a function of height for $\theta = 3\pi / 8$, corresponding to the dotted line in panels $a$ and $b$.}
	\label{fig: continua-thetavary}
\end{figure}

It should be noted that, contrary to Figure \ref{fig: nye-thomas-continua}, the continua show sharp transitions and spiked behavior which is an unavoidable consequence of the cooling curve used. In the case of Figure \ref{fig: nye-thomas-continua} the continua are smooth because the equilibrium temperature was uniform, such that we only needed to evaluate the cooling curve and its derivatives in a single value. Here however we have a wide range in temperatures, and while the heat-loss function $\HL$ itself is continuous, the cooling curve has discontinuous derivatives since we do a local second-order (linear) approximation in between two data points (that is, between two `original' table values) in order to preserve the original table as close as possible. As a consequence of this it is entirely possible that for some spatially varying temperature profiles there may be jumps in the continuum curves near temperature transitions between two table points. If one were to use a different representation for the cooling curve, that is, one that also has continuous derivatives $\HL_T$ and $\HL_\rho$, this would yield smooth curves in the complex eigenvalue plane.

Figure \ref{fig: continua-thetavary} keeps the magnitude of $k$ equal but varies the angle $\theta$ between $\bfk$ and the magnetic field, hence $k_y = k_0\sin\theta$ and $k_z = k_0\cos\theta$. Stability regions for the thermal and slow continua are shown for $k_0 = 1$ and $\theta$ between $0$ and $2\pi$, the case with $\theta = 0$ corresponds to the dotted line case at $k_z = 1$ in Figure \ref{fig: continua-kzvary}. As the wave vector becomes more perpendicular to the magnetic field the thermal continuum becomes more and more unstable, due to the fact that parallel thermal conduction is less efficient for larger angles. For near perpendicular propagation the thermal continuum is unstable over the entire domain, owing to the almost complete loss of stability provided by parallel conduction. The slow continuum is mostly stable, with a few unstable regions similar to the previous case.
In this Figure \ref{fig: continua-thetavary}, panels $c$ and $d$ show the continua as a function of height, corresponding to the dotted horizontal line at $\theta = 3\pi / 8$.
We conclude that in the framework of single-fluid MHD adopted here, the entire chromosphere to corona of our realistic model is found liable to thermal instability, while the lower coronal regions in addition demonstrate slow continuum overstabilities, with corresponding traveling, singular modes at altitudes between $\approx 2.5 - 8$ Mm.

\subsection{Full MHD spectra for the solar corona}
We now purely focus on the solar corona, limiting the lower boundary of the slab to just after the transition region at $x = 2.5$ Mm, with the upper boundary set to $x = 25$ Mm.
The magnetic field is still aligned along the $z$-axis with a magnitude of $10$ Gauss, the wave numbers are chosen to be $k_y = 0$ and $k_z = 0.8$ in normalized units. The latter corresponds to a wavelength of $\approx 7.8$ Mm, compared to a slabsize of $22.5$ Mm thickness. Based on Figure \ref{fig: continua-kzvary} this will bring the equilibrium into the unstable thermal and slow continuum regions.

\begin{figure}[t]
	\centering
	\includegraphics[width=\textwidth]{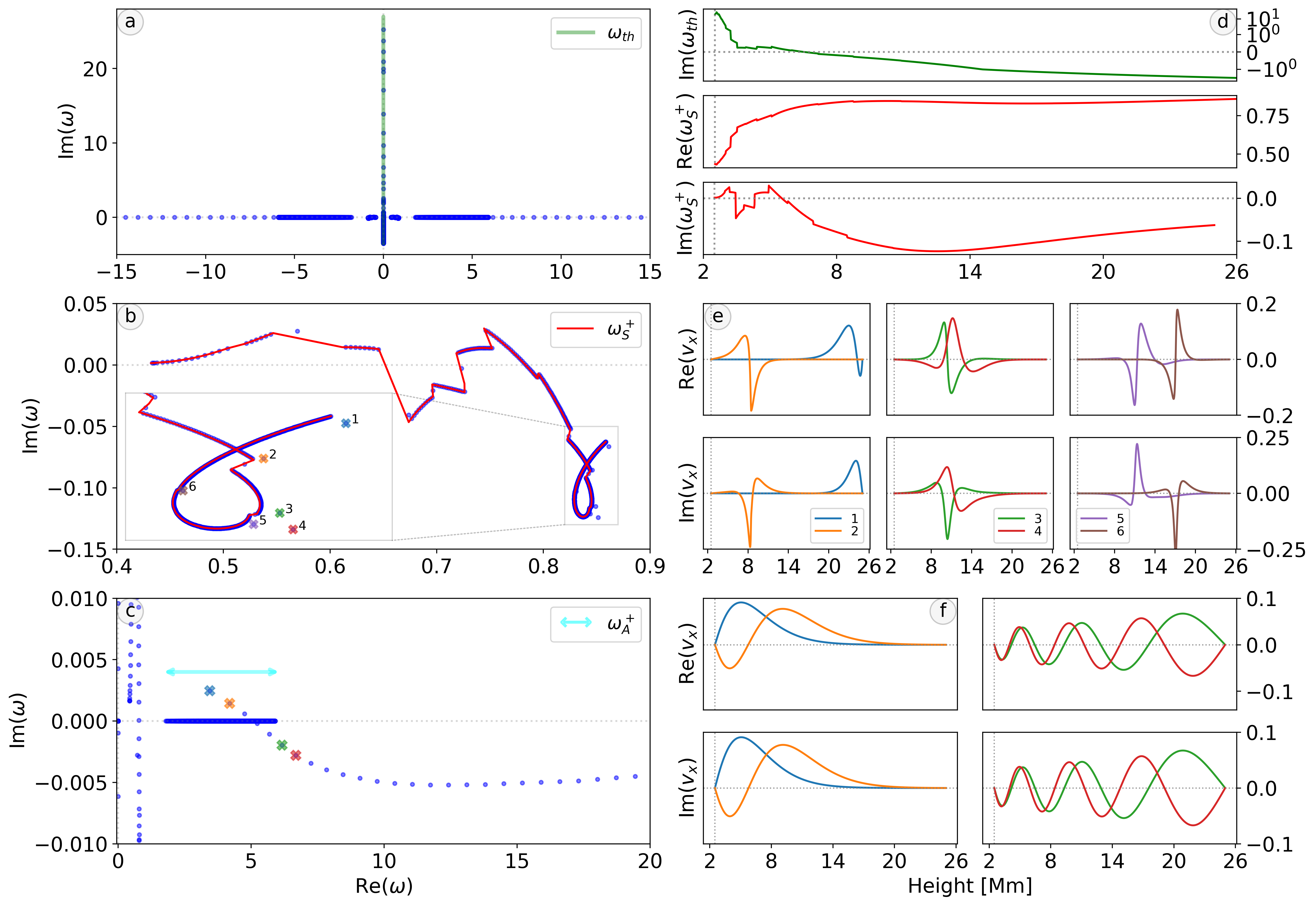}
	\caption{Spectrum with eigenfunctions of a solar coronal slab, based on the profiles in Fig. \ref{fig: profile} for $x \in [2.5, 25]$ Mm, $B_{0z} = 10$ G and $k_z = 0.8$. Panel $a$: part of the full spectrum with focus on the unstable thermal continuum.
		Panel $b$: zoom-in near the slow continuum, the inset focuses on the set of discrete stable slow modes. Panel $c$: zoom of panel $a$, showing the Alfv\'en and partially unstable fast sequences. Panels $d$: thermal and slow continua as a function
		of height. Panels $e$: $v_x$ eigenfunctions of the six discrete slow modes marked in panel $b$, inset. Panels $f$: $v_x$ eigenfunctions of the four fast modes marked in panel $c$. All eigenfunctions are normalized.}
	\label{fig: corona-parallel}
\end{figure}

\begin{figure}[t]
	\centering
	\includegraphics[width=\textwidth]{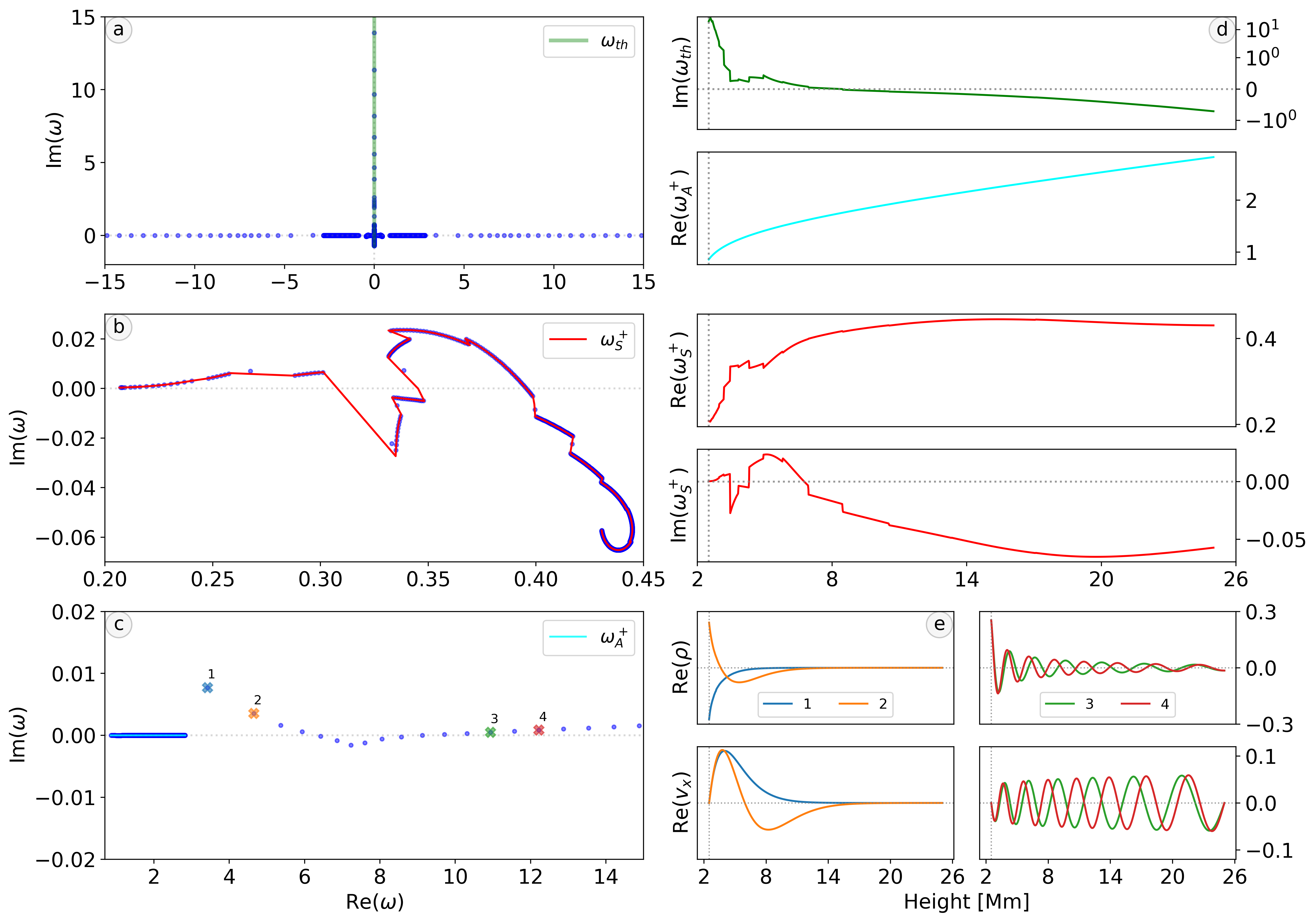}
	\caption{Spectrum with eigenfunctions of a solar coronal slab for $x \in [2.5, 25]$ Mm, $B_{0z} = 10$ G, $k_0 = 1$ and an angle $\theta = 3\pi / 8$ between $\bfk$ and $\bfb_0$. Panel $a$: part of the full spectrum with focus on the unstable thermal continuum.
		Panel $b$: zoom-in near the slow continuum. Panel $c$: focus on the Alfv\'en continuum and unstable fast branch with density and velocity eigenfunctions in panels $e$. Panels $d$: thermal, Alfv\'en and slow continua as a function of height.
		All eigenfunctions are normalized.}
	\label{fig: corona-notparallel}
\end{figure}

Figure \ref{fig: corona-parallel} shows the spectrum with corresponding eigenfunctions of such a coronal slab, obtained through \texttt{Legolas} using 500 grid points. Panel $a$ focuses mostly on the unstable thermal modes, with the continuum range annotated by a green line. As expected from Fig. \ref{fig: continua-kzvary} this has both a stable and unstable part, with growth rates approximately two orders of magnitude larger than the unstable modes in the fast or slow sequences. The two continua are given as a function of height in Fig. \ref{fig: corona-parallel}, panels $d$, with the thermal continuum shown in the top panel and the real and imaginary parts of the slow continuum shown in the second and third panels on the right, respectively.
Panel $b$ zooms in near the slow continuum, annotated in red, tracing out an intricate profile in the complex eigenvalue plane with various discontinuities due to the jumps in the derivatives of the cooling curve.
The slow continuum is partially unstable and loops back on itself, revealing six discrete damped slow modes, annotated with numbers $1$ through $6$. Their $v_x$ eigenfunctions are given in panels $e$. No other discrete modes were found; the seemingly discrete unstable slow mode near Re$(\omega) \approx 0.57$ is actually a continuum mode, characterized by a $1/x$ behavior in its eigenfunctions. This mode is essentially a remnant of a discontinuity in the continuum present there, since there are no modes underneath the solid red line.

Technically speaking the continuum in this case consists of various segments, which is immediately clear from the regions of clustered eigenvalues in the continuum range. The gaps between these segments correspond to the various discontinuous jumps in the continuum.
Panel $c$ shows the (purely real) Alfv\'en continuum, the range of which is annotated with cyan arrows, together with the first few modes of the fast sequence which accumulates at infinity. The first three fast modes are overstable, then the sequence becomes (and remains) damped. $v_x$ eigenfunctions of the first two overstable modes and two damped modes are given in panel $f$. All eigenfunctions are normalized and color-coded to their annotations in the corresponding panels. Hence, this particular full spectrum at prechosen wavenumber for the solar corona demonstrates unstable thermal and slow continua, as well as the possibility of overstable fast modes in the Alfv\'en continuum range (which in this case are uncoupled with it due to $k_y = 0$).

Next we look at a more general case, where the wave vector $\bfk$ is no longer parallel to $\bfb_0$. Based on our continuum analysis this will lead to configurations that are more unstable due to the partial loss of stabilization provided by parallel thermal conduction. Here we take the magnitude of $\bfk$ equal to unity and take an angle $\theta = 3\pi / 8$ between the wave vector and the magnetic field, corresponding to the horizontal dashed line in Fig. \ref{fig: continua-thetavary}. The spectrum (500 grid points) is shown in Figure \ref{fig: corona-notparallel}. On large scales (panel $a$) this looks quite similar to the parallel case: the thermal continuum is again mostly unstable with dominating growth rates compared to the other mode sequences. Panel $b$ shows the slow continuum, which in this case does not loop back on itself and has no discrete modes associated with it. The thermal, Alfv\'en and slow continua are shown as a function of height in panels $d$ from top to bottom, respectively.
Panel $c$ zooms in on the Alfv\'en continuum (cyan line) and fast mode sequence, with no overlap between the various eigenvalues. A few overstable fast mode overtones are annotated with their density and $v_x$ eigenfunctions shown in panels $e$. Contrary to the parallel case the fast branch crosses the marginal stability line $\omega_I = 0$ again after a few stable modes, hence most of the fast sequence of discrete modes here are overstable. Note that at this angle, the fast sequence shows no overlap with the Alfv\'en continuum, so no resonant behavior is at play, and their overstability is entirely due to the non-adiabatic effects (thermal conduction and optically thin radiative losses) included.

\subsection{MHD spectra for the solar chromosphere}
Lastly we shift our focus to the chromosphere, as such we take our slab between $x = 0$ and $x = 2.1$ Mm which ends just before the transition region (see Fig. \ref{fig: profile}). The wave vector $\bfk$ is taken to be parallel to the magnetic field. The choice of magnetic field strength will strongly influence the plasma-$\beta$ magnitude, including the height at which the crossing $\beta > 1$ to $\beta < 1$ occurs, which in turn has a major effect on the behavior of the slow continua. Hence, we will perform a magnetic field strength study in this Subsection, and look how the varying magnetic field affects the spectra (and continua) as a whole. The slabsize will remain the same in all cases.

\begin{figure}[t]
	\centering
	\includegraphics[width=\columnwidth]{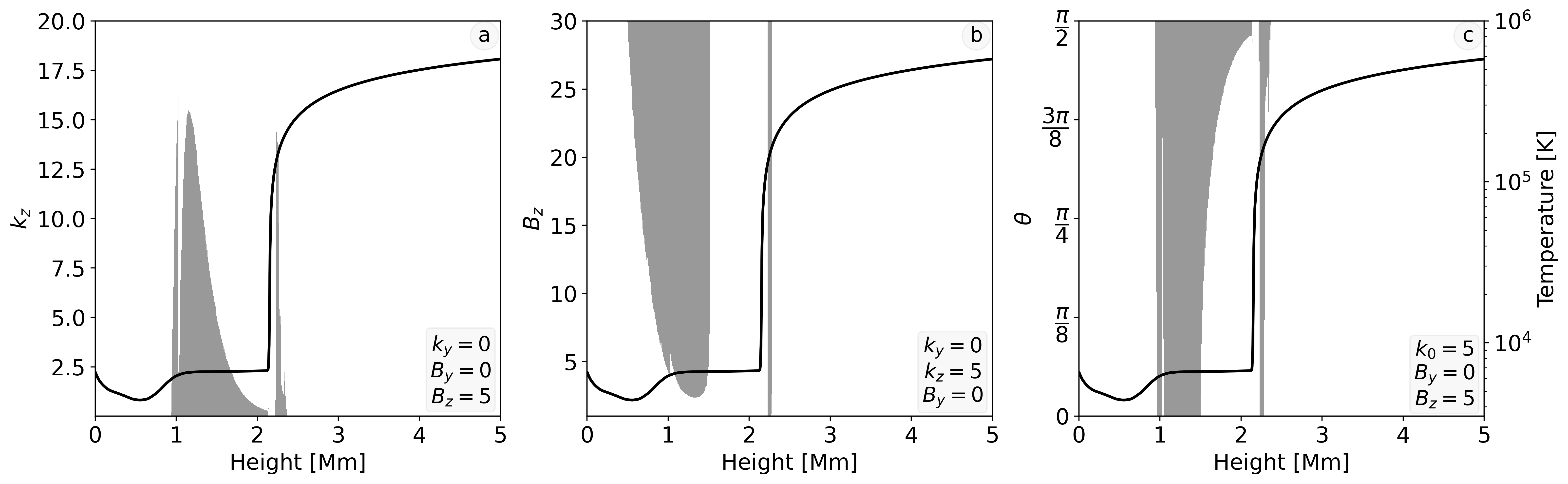}
	\caption{
		Plots of Equation \eqref{eq: imaginary_solutions} for varying $k_z$ (panel $a$), $B_z$ (panel $b$) and angle $\theta$ between $\bfk$ and $\bfb_0$ (panel $c$). Shaded regions indicate conditions where the continuum equation \eqref{eq: continuum_modes}
		has three purely imaginary solutions. The black line denotes the temperature profile, all quantities are normalized to a length scale of $1$ Mm, magnetic field of $10$ Gauss and temperature of $1$ MK.
	}
	\label{fig: imaginary_solutions}
\end{figure}

Before looking into the actual spectra themselves it is worth having a short interlude on a very special case of the continuum equation \eqref{eq: continuum_modes}. In most cases, including the ones discussed up to now, the solutions of this equation correspond to one purely imaginary solution (the thermal continuum) and two complex conjugate solutions (the slow continua). However, since this equation is essentially just a third-order polynomial, in some cases the possibility exists to have three purely imaginary solutions, which means that the slow continua lay \textit{on} the imaginary axis. If this occurs, then it is a priori difficult to say which one of those three imaginary modes belongs to the thermal continuum, and which ones belong to the forward (i.e. positive) or backward (i.e. negative) slow continua. The criterion for a third order polynomial to have three completely imaginary solutions is defined purely by a function of its coefficients, and as such this can be analytically determined. It can be shown that the continuum equation \eqref{eq: continuum_modes} will have three completely imaginary solutions if the following condition is satisfied:
\begin{equation} \label{eq: imaginary_solutions}
	\begin{gathered}
		(\gamma - 1)^2 \left\{2 (\gamma - 1)^2 \Qai^3 + 9 c^2_A k_\parallel^2 \rho_0^2 \left(c^2_A + c^2_s\right) \left[3 \Qi \left(c^2_A + c^2_s\right) - \Qai c^2_s\right]\right\}^2 \\
		- 4 \left[(\gamma - 1)^2 \Qai^2 - 3 c^2_A c^2_s k_{\parallel}^2 \rho_0^2 \left(c^2_A + c^2_s\right)\right]^3 \leq 0,
	\end{gathered}
\end{equation}
where we defined
\begin{equation}
	\Qi = \Qifull \qquad \text{and} \qquad \Qai = \Qaifull
\end{equation}
for brevity. Figure \ref{fig: imaginary_solutions} shows parametric plots of Equation \eqref{eq: imaginary_solutions}, where we used our solar atmosphere model and varied the wave number $k_z$ (panel $a$), the magnetic field component $B_z$ (panel $b$) and the angle
between the $\bfk$-vector and $\bfb_0$ (panel $c$) in the chromosphere and lower corona. The shaded regions indicate conditions where Eq. \eqref{eq: imaginary_solutions} is satisfied and where we thus have three purely imaginary solutions to the continuum equation. The temperature profile is overlayed with a black line.

Note that in adiabatic MHD there is no confusion between the forward and backward slow continua possible, and since our equilibrium is static (no flow), the two wavelike solutions then lie symmetric with respect to the marginal frequency on the real frequency axis. In non-adiabatic MHD, the common situation for the slow continuum is to have two complex conjugate solutions, still symmetric against mirror reflection of the imaginary Re$(\omega) = 0$ axis. Now, the possibility exists for these to merge into purely non-propagating modes, and these ``ascend" and ``descend" on the imaginary axis as we vary $x$. Together with the thermal mode, then three such non-propagating modes may exist for each $x$-location, and their mutual ordering is \textit{not} fixed.

This can be related to work by \citet{waters2019} in hydrodynamics (see their Table 1), where they investigate mode stability of acoustic waves and thermal modes, and also encounter certain regions where all of these are purely imaginary. These authors argue then to relabel these modes as E, S and F for entropy, slow and fast isochoric modes. However, relabeling the wave modes (acoustic waves and the entropy mode) based on their growth rates (that is, based on $\omega_I$) does not make much sense. This is especially true for MHD, where we have seven wave modes instead of three. The core of (ideal) MHD spectral theory is the existence of slow and Alfv\'en continua, together with the fast accumulation point at infinite real frequency, and this is here extended to non-adiabatic settings.
It is entirely possible that scenarios exist where the thermal mode is stable, for example, but lies in between or below the imaginary slow modes on the imaginary axis.
Hence, one should be careful to state that the mode with highest imaginary growth rate is always the thermal mode, since this is not guaranteed to be the case. This will ultimately be reflected in the mode eigenfunctions and their polarizations. These eigenfunctions are singular for all continuum modes.

\begin{figure}[t]
	\centering
	\includegraphics[width=\textwidth]{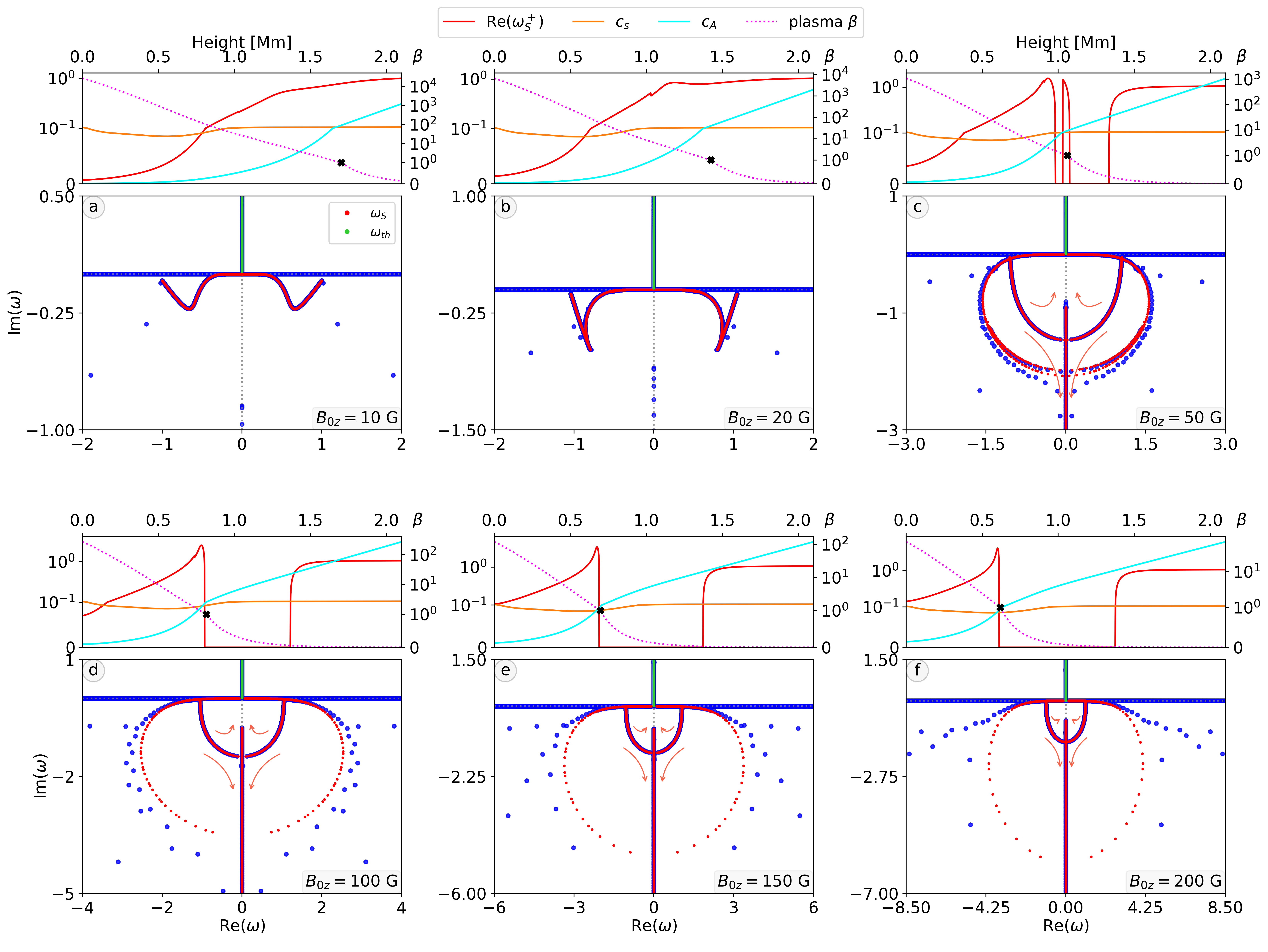}
	\caption{Partial spectra and profiles for six different (uniform) magnetic field strengths in the solar chromosphere with $x$ between 0 and $2.1$ Mm, with focus on the thermal and slow continua.
		The wave vector is parallel to $\bfb_0$, aligned with the $z$-axis and $k_z = 10$.
		The value for $B_z$ is annotated in the bottom-right corner of every spectrum, the thermal and slow continua are indicated with green and red dots, respectively. The top figure of every panel shows the real part of the slow continuum, the sound speed
		and the Alfv\'en speed as a function of height in red, orange and cyan, respectively, with normalized values given on the left $y$-axis. Plasma $\beta$ is given in a magenta dashed line with values on the right $y$-axis, the black cross denotes the
		point where $\beta = 1$.}
	\label{fig: chromosphere}
\end{figure}

As it turns out (see Figure \ref{fig: imaginary_solutions}), interesting slow continuum behavior actually occurs in the solar chromosphere when the magnetic field crosses a certain threshold.
Figure \ref{fig: chromosphere} contains six panels labeled $a$ through $f$, all of which contain two subpanels. The top one shows the sound speed in orange, Alfv\'en speed in cyan and real part of the slow continuum in red as a function of height. All these quantities are normalized and have their values associated with the left $y$-axis. The magenta dashed line depicts the plasma-$\beta$ as a function of height with values given on the right $y$-axis, the black cross annotates the $\beta = 1$ transition point. The bottom subpanel of each panel zooms in on the slow continua, which are both (that is, $\omega_S^+$ and $\omega_S^-$) shown with red dots. The thermal continuum is annotated with green dots and is completely unstable in all cases: no thermal modes will cross the stability line $\omega_I = 0$ into the stable regime. Every one of the six panels has a different magnetic field strength, labeled in the bottom-right corner of each spectrum figure. Each of these spectra is calculated using $500$ gridpoints and in all cases the wave vector is parallel to the magnetic field, both aligned with the $z$-axis.
In order to have a comparable ratio between wave length and slabsize as Fig. \ref{fig: corona-parallel}, we take $k_z = 10$ in normalized units corresponding to a wavelength of $\approx 630$ km compared to a slabsize of $2100$ km.

Panels $a$ and $b$ correspond to weak magnetic field strengths of $10$ and $20$ Gauss, respectively, corresponding to plasma-$\beta$ values of $\approx$$10^4$ at $x = 0$. The thermal continuum is unstable, the slow continua are fully stable. When doubling the magnetic field from $10$ G to $20$ G (panels $a$ to $b$), the slow continuum starts moving downwards into the stable part of the complex plane and exhibits a sharp peak near the bottom, indicating a downwards trend for increasing magnetic field. In panel $c$ we crossed the threshold (which occurs at approximately $B_{0z} \approx 35$ G) where certain parts of the slow continuum have collapsed to purely imaginary solutions. Now the plasma beta at $x=0$ is approximately $\beta \approx 10^3$. The initial downwards arches visible in panel $b$ led to a connection, where part of the slow continua continues upwards and part continues downwards on the imaginary axis as indicated by the arrows. Again, we should stress that it is impossible to know if it is the right side or left side branches that continue upwards (or both). In this case we do know that both purely damped eigenmodes are slow continuum modes, since the thermal continuum is unstable.

When the magnetic field keeps increasing this trend continues, as the outer loop continues further outwards and downwards (panels $d$ through $f$) with plasma-$\beta$ values at $x=0$ decreasing from approximately 270 to 70. Interestingly, it has proven extremely difficult to properly resolve the outer continuum loop. All these spectra were calculated at 500 grid points, although even at a resolution of 1500 grid points \texttt{Legolas} manages to resolve the inner continuum loop, while the outer loop starts out resolved near the $\omega_R = 0$ axis and then moves further away. This suggests that extreme resolutions are required to actually probe those outer continuum modes which is computationally challenging, at least with the currently available eigenvalue solvers in \texttt{Legolas}.
It is also worthwhile to note that the continuum collapse seems to occur at or at least near the point where the sound speed $c_s$ equals the Alfv\'en speed $c_A$ and the transition $\beta = 1$. The reason this does not occur in panels $a$ and $b$ is due to the fact that the magnetic field is too weak to influence the slow continuum strongly, and $B_{0z} \approx 35$ G is the turning point for this particular configuration.

In conclusion, we showed that the MHD magnetothermal spectrum is interesting in its own right for quantifying thermal stability in the chromosphere. Our results show that thermal modes are almost always unstable, while slow continua may turn into purely non-propagating eigenmodes that can in principle also become unstable. Note however that this MHD eigenspectrum of the chromosphere adopted the same optically thin radiative losses as shown in Figure \ref{fig: coolingcurve} for computing its eigenspectrum.
The actual chromosphere is partially ionized and although this is partially accounted for in the cooling curve through the low temperature extension from \citet{dalgarno1972}, some of our single-fluid, perfectly ionized MHD model assumptions may need to be revisited.

\section{Conclusion} \label{sec: conclusion}
In this paper we introduced a detailed spectroscopic analysis of both analytical and more realistic, numerical solar atmosphere models. We started out by revisiting an analytical model of the solar atmosphere, based on work by \citet{nye1976}. We showed that we can reproduce the spectra discussed in the original work to high accuracy, however, the analytical treatment of this model is only handling discrete fast modes, and only feasible for wave vectors parallel to the magnetic field since this goes accompanied by a decoupling of the Alfv\'en continuum. Using \texttt{Legolas} however, we extended the work in \citet{nye1976} by first looking at a slight deviation from parallel propagation, where we showed that this has not much effect on the spectrum itself but introduces locally singular behavior in the eigenfunctions related to resonance between the fast modes and the Alfv\'en continuum. This was then extended even further by including non-adiabatic effects, which had a profound influence on the spectrum and the slow continua, revealing unstable thermal, slow and fast modes.

Next we moved on to a fully realistic solar atmosphere model, obtained by numerically integrating a tabulated temperature model from \citet{avrett2008}. A detailed continuum analysis revealed that for this particular model most of the solar chromosphere is thermally unstable and is hence quite susceptible to thermal instability. The solar corona on the other hand was revealed to be more stable for both the thermal and slow continua, with decreased stability for wave vectors more perpendicular to the magnetic field due to less efficient parallel thermal conduction. The actual spectrum for the solar corona proved to be quite complex, with the slow continuum tracing out an intricate pattern in the complex eigenvalue plane having both stable and unstable regions. In all cases however the thermal continuum was shown to be the most unstable, with growth rates almost two orders of magnitude larger than the most unstable modes of the slow continuum or fast sequence. It is to be emphasized that the model adopted for the solar corona is extremely realistic, and shows that linear theory can in a sense predict where condensations may likely be forming in follow-up nonlinear simulations.

For the solar chromosphere we encountered interesting behavior for the slow and thermal continua, where we came across regions where the solutions to the continuum equation in \eqref{eq: continuum_modes} collapsed to three purely imaginary solutions. The effect of this is that the slow continua come together on the $\omega_R = 0$ line, with modes then moving upwards or downwards on the imaginary axis.
Physically speaking this has a couple of interesting consequences. Based on Figure \ref{fig: chromosphere} the conclusion can be drawn that most of the solar chromosphere is quite susceptible to thermal instability. However, if the collapse of the slow continuum occurs frequently in the chromosphere, situations may arise where the upwards branch of slow continuum modes on the imaginary axis crosses the marginal stability line $\omega_I = 0$, becoming unstable. In that case it will depend on how strongly the growth rates of the thermal continuum are affected by this collapse, and whether or not they become less stable/unstable compared to the imaginary slow continuum modes. All of this will of course depend on the magnetic field strength and additional effects like perpendicular thermal conduction or ambipolar diffusion, to name a few.

Generally speaking, in most cases we are dealing with an unstable thermal continuum accompanied by intricate behavior in the stable part of the complex plane, so physically speaking these stable modes would damp out rather quickly and the only ``dominant" behavior would be due to the unstable thermal modes, leading to thermal instability and mostly in-situ condensations. However, in some cases even the slow and fast modes can become unstable, and if this happens then it is a matter of which growth rate is dominant. This holds especially true if the continuum solutions collapse to the imaginary axis, where a subsequent shift in mode solutions can make a given configuration suddenly stable or unstable. An actual initial value-boundary value simulation of the magnetized solar atmosphere would basically relate to the spectra shown here, in the sense that the then required Laplace-transform would pick up the unstable and overstable parts of the linear MHD spectrum (see \citet[chapter 10]{book_MHD}).

Note that we did not include perpendicular thermal conduction in any of the cases treated in this work, for the straightforward reason that inclusion of this effect is a research topic in itself. Including a non-zero $\kappa_\perp$ will replace the thermal continuum with a quasi-continuum, which is a dense band of discrete thermal modes \citep{vanderlinden1991}. This has profound influence on the actual spectrum and eigenfunctions and introduces an additional stabilizing effect on the thermal modes. In this context the discussion on unstable slow modes and the collapse of the continuum solutions would become even more relevant, since this additional stabilizing effect may switch the dominating behavior from the thermal mode to an unstable slow mode.

To conclude, all cases studied had a simple uniform horizontal magnetic field, so without shear, currents, or vertical magnetic field components. Even this simple magnetic field configuration already reveals just how complex the quantification of actual eigenmodes of a true solar atmosphere can be. Having detailed information on the behavior of the continua already goes a long way in predicting (in)stability of any given equilibrium state. Since thermal instability lies at the very basis of prominence formation and coronal rain, we show here just how easy it is to have thermally unstable regions. Physically, in these regions any sufficiently large perturbation will without a doubt give rise to thermal instability and hence possible prominence formation. Exactly \textit{how} this occurs however will depend on the intricacies of the spectrum, whether it is an unstable slow mode, fast mode or thermal mode, or a coalescence of unstable modes. That the dominating mode of the spectrum at any given point mostly governs the dynamical temporal evolution of the system, could be verified by follow-up nonlinear, non-adiabatic MHD numerical simulations.




\begin{acks}
	This work is supported by funding from the European Research Council (ERC) under the European Unions Horizon 2020 research and innovation programme, Grant agreement No. 833251 PROMINENT ERC-ADG 2018;
	by the VSC (Flemish Supercomputer Center), funded by the Research Foundation – Flanders (FWO) and the Flemish Government – department EWI; and by internal funds KU Leuven, project C14/19/089 TRACESpace.
\end{acks}



\bibliographystyle{spr-mp-sola}
\bibliography{bibfile}

\end{article}
\end{document}